\documentclass[aps,prd,amsmath,amssymb,superscriptaddress,preprintnumbers,preprint, nofootinbib]{revtex4-1}
\pdfoutput=1
\usepackage{multirow}
\usepackage{amssymb}
\usepackage{amsmath}

\usepackage{braket}
\usepackage[usenames,dvipsnames]{color}  
\usepackage{array}              		                 		
\usepackage{graphicx}	
\usepackage{tikz}		
\usepackage{adjustbox}		
\usepackage{setspace}
\usepackage{tabstackengine}
\stackMath

\usepackage{longtable}

\usepackage[linktoc=none]{hyperref}
\usepackage[normalem]{ulem}
\usepackage{dsfont}
\usepackage{enumitem} 
\makeatletter
\def\p@subsection{}
\makeatother

\definecolor{darkred}{rgb}{0.6,0,0}

\definecolor{linkcolor}{rgb}{0,0,0.5}

\def\Z2{$\mathcal{Z_2}$}

\def\321{$\mathrm{SU(3) \otimes SU(2) \otimes U(1)}$ }

\newcommand{\AddrUnina}{Dipartimento di Fisica E. Pancini, Universit\`a di Napoli Federico II \\ Complesso Universitario di Monte Sant'Angelo, Via Cintia, Napoli (NA), Italy \\ and INFN, Sezione di Napoli.}
\newcommand{\AddrSalento}{Dipartimento di Matematica e Fisica E. De Giorgi, Universit\`a del Salento, \\ via per Arnesano 73100 Lecce (LE), Italy \\ and INFN, Sezione di Lecce.}

\begin{document}
  
\title{
331 Models and Bilepton Searches at LHC }
\author{Roberta Calabrese}\email{rcalabrese@na.infn.it}
\affiliation{\AddrUnina}
\author{Alberto Orso Maria Iorio}\email{albertoorsomaria.iorio@unina.it}
\affiliation{\AddrUnina}
\author{Stefano Morisi}\email{smorisi@na.infn.it}
\affiliation{\AddrUnina}
\author{Giulia Ricciardi}\email{giulia.ricciardi2@unina.it}
\affiliation{\AddrUnina}
\author{Natascia Vignaroli}\email{natascia.vignaroli@unisalento.it}
\affiliation{\AddrSalento}\affiliation{\AddrUnina}

 \begin{abstract}

Despite being remarkable predictive, the Standard Model  leaves unanswered several important issues, which motivate an ongoing search for its extensions. 
One fashionable possibility are the so-called 331 models, where the electroweak gauge group is extended to  $SU_L(3)\times U(1)$. 
We focus on a minimal extension which includes vector-like quarks (VLQs) and new gauge bosons,  performing a consistent analysis of the  production at  LHC of a pair of doubly-charged bileptons. We include for the first time all the relevant processes where  VLQs contribute, and in particular the associate production VLQ-bilepton. 
Finally, we extract the bound on the bilepton mass, $m_Y>$ 1300 GeV, from a reinterpretation of a recent ATLAS search for doubly-charged Higgs bosons in multi-lepton final states. 
   
\end{abstract}
\maketitle

\section{Introduction}

The Standard Model (SM) based on the group $G_{\rm SM}= SU(3)_c\times G_{\rm ew}$,  where  $G_{\rm ew}=SU(2)_L\times U(1)$, is a remarkably successful theory, but it has been proven to be  incomplete,
and 
needs to be extended.
Since the three gauge couplings, running with  energy,  are drawn nearer around $10^{15}$ GeV, an appealing possibility is that
$G_{\rm SM}$ arises from the spontaneously breaking of a larger gauge group, characterized by only one structure constant  which
unifies strong and electroweak interactions.
Most Grand Unified Theories (GUTs) are based 
on the gauge groups $SU(5)$ and $SO(10)$\,\cite{Ross:1985ai}.
They
predict proton decay, whose observation in itself provides
a strong motivation. 
However so far experiments  are just putting stronger and stronger lower limits on proton life-time.

A completely different approach is to consider minimal extensions of 
$G_{\rm SM}$ with not-simple gauge groups. 
There are 
 many possibilities. The simplest one is to enlarge the electroweak sector to $G_{\rm ew}\times U(1)'$ to add a non-standard
neutral gauge boson generally named $Z'$, see for instance Ref.
\cite{Langacker:2008yv}. The next case is inspired by the request to symmetrise the SM matter assignment, whose complete asymmetry between left and right-handed fields is probably  one of its less appealing feature.
In the so called Left-Right models,
$G_{\rm ew}$ is embedded into $SU_L(2)\times SU_R(2)\times U(1)$\,\cite{Mohapatra:1974hk, Senjanovic:1975rk}. 
Another interesting possibility 
is to extend $SU_L(2)$ with $SU_L(N)$. 
In Refs. \cite{Schechter:1973nqg, Gupta:1973pv, Albright:1974nd, Georgi:1978bv},
$G_{\rm ew}$ is extended to $SU_L(3)\times U(1)$. These seminal works focus on the possibility to extend the GIM mechanism \cite{Glashow:1970gm} 
to more than 4 quarks and to
suppress unwanted flavor changing neutral currents. Then in \cite{Singer:1980sw}
it has been proposed the first 
$SU_L(3)\times U_X(1)$ (where $X$ is a generalized hypercharge, see below) model where the three families of fermions are embedded into irreducible representations of the gauge group. 
The first complete model based on $SU_L(3)\times U_X(1)$ (hereafter named $331$ where the first 3 refers to $SU_c(3)$)
has been proposed by Pisano, Pleitez \cite{Pisano:1992bxx} and Frampton \cite{Frampton:1992wt}. 

In 331 models
 the hypercharge operator ${\hat Y}$ is given by
\begin{equation}
	\frac{\hat Y}{2} =\beta \hat T^8+X{\mathds 1} \; ,
\label{hyper:ID}
\end{equation}
where   $X$ is the quantum number associated with $U_X(1)$ and $\beta$ is a  free parameter.
Any 331 model is characterized in general by the pair $(\beta,X)$, whose values are obviously related  to the  electric charges of the fields (fermions and bosons). By fixing the charges of the standard model fermions, the $X$-charges  
are expressed as a function of the parameter $\beta$. Therefore 331 models can be classified in terms of the value of $\beta$. The assumed correlation SM charges-$\beta$ value does not naturally hold in cases where one considers additional multiplets including only BSM  fermions, which  then may  have different $X$ values given the same $\beta$ (see e.g. Ref. \cite{Descotes-Genon:2017ptp}). \\
In this work we consider  a minimal requirement that the new gauge bosons have at least integer charge values, although there are 331 models that discard this assumption, see for instance  \cite{RomeroAbad:2020zpv}. This implies that $|\beta|=n/\sqrt{3}$, where $n$ is an odd integer number. On the other hand, by requiring that all gauge couplings are real, it follows that 
$ |\beta | <  1.83 $ (see also \cite{Fonseca:2016tbn}). Therefore, models with $n>3$ are excluded and the only possible cases are $\beta=\pm1/\sqrt{3},\pm \sqrt{3}$.

Besides the specific value of the parameter $\beta$, 331 models are  classified in terms of matter assignment into irreducible representation of $SU_L(3)$, namely triplets and anti-triplets.
It is well known that the anomaly cancellation implies that quarks families cannot all be assigned  to triplets or anti-triplets. We derive the important relation
$$N_Q^{(3)} =  ( 2 N_F - N_\ell^{(3)} ) /3 $$ 
where $N_{Q,\ell}^{(3)}$ is the number of quark (lepton) families transforming as $SU_L(3)$-triplet and $N_F$ is the number of families.
Both sides of the equality must be integers. This is  true if the number of families $N_F$ is a multiple of the color $N_c=3$, as in minimal models where $N_\ell^{(\bar3)}=3$, $N_Q^{(3)}=2$ and $N_Q^{(\bar3)}=1$. However, such proportionality is not strictly necessary, contrary to what is typically reported in literature. 
An example is  $N_F=4$ with $N_\ell^{(3)}=N_\ell^{(\bar{3})}=2$ that gives $N_Q^{(3)}=N_Q^{(\bar{3})}=2$ (see also Ref.\,\cite{Diaz:2004fs}).

Another constraint on 331 models is given by $SU(3)_c$ asymptotic freedom. As in SM QCD, perturbative calculations limit the number of quark flavours to be  equal to or less
than 16. Since each family in 331 models has three flavours, that yields a constraint on the number of families: $N_F\le 5$.

More details, including a general  classification of 331 models and the passages needed to engineer a 331 model compatible with the SM, i.e. the choice of the scalar sector and the  charge assignment of the matter fields,  will be found in Appendix.

Even if from the group theory  point of  view extending $SU_L(2)$ to $SU_L(3)$ represents a small modification, from the phenomenological point of view it has important consequences:
\begin{itemize}
    \item The matter assignment requires the introduction of
    extra fermions that can have exotic charges and can be vector-like. In the minimal model we consider there are two vector-like quarks with charges $-4/3$ and $5/3$.
    \item The scalar sector must be extended with extra Higgs fields. For example, in our case the model contains three scalar $SU_L(3)$ triplets.
    \item The model has 5 more gauge bosons that can have exotic charges. In our case the extra gauge bosons are $Z'$, $W'^{\pm}$ and the doubly-charged bilepton $Y^{\pm\pm}$. 
\end{itemize}
The wealth of 331 models of new physics yields to opportunities for searching at colliders. 
We provide an inclusive study of the multi-lepton channel $pp\to \ell^{\pm}\ell^{\pm}\ell^{\mp}\ell^{\mp} +X$ at LHC mediated by 331 new fields, in particular by the pair production of bileptons $Y^{\pm\pm}$. The multi-lepton production process cross section is sizable modified by the presence of 331 new physics. 
This analysis tests a general version of 331 models with $\beta=\sqrt{3}$ and a minimal set of assumptions:
\begin{itemize} \item all matter multiplets include SM fermions, excluding the possibility of fermion multiplets including only BSM fields;
\item all gauge bosons, including BSM ones,  have integer charges;
\item all right-handed, $SU(3)$ singlet, fermions have a left-handed counterpart.
\end{itemize}

The paper is organized as follows:
in section \ref{sec:model} we present the main features of the model and in section \ref{sec:pheno} we discuss its phenomenology. The analysis of bilepton production at LHC is reported in section \ref{sec:bilepton_prod}, and conclusions are given in section \ref{sec:conclusions}.

\section{The model}\label{sec:model}

\subsection{Matter assignment}

In this work, we focus on the  minimal 331 model \cite{Corcella:2018eib}, where the first two generations of quarks transform as triplets under $SU(3)_L$, whereas the third generation and all the leptons transform as anti-triplets.    In the appendices we discuss a general classification of 331 models, and provide 
a detailed view of the one we consider. In the classification scheme presented in the appendices this model  is indicated as  $A[\sqrt{3}]$.
Table\,\eqref{tab:fermion} contains a summary of the matter content of the model. 
The fermions $D^{1,2}$ and $T$ in the color triplet are exotic quarks with fractional electric charges $-4/3$ and $5/3$ respectively \footnote{These exotic quarks do not have  the same value charge of down type quarks and top quark, but maintain  the same charge sign. 
}, while the fermions  $\zeta_L^a$, $a$ = 1, 2, 3, which are color singlets, are leptons with positive charge $+1$. 
All three BSM quarks can be assumed to carry the usual baryonic number for quarks, namely $B=1/3$.
In the spirit of minimally extending the SM, we identify $\zeta^a_L = (\ell^a_R)^c$. 

\begin{table}[h]
\centering
\begin{tabular}{|c|ccc|c|ccc|}
    \hline
    \hline
& $SU(3)_c$ & $SU(3)_L$     & $U(1)_X$ & &  $SU(3)_c$ & $SU(3)_L$      & $U(1)_X$                                   \\ 
\hline
\multirow{3}{*}{$\left(\begin{array}{c}u^{1,\,2}_L\\ d^{1,\,2}_L\\ D^{1,\,2}_L\end{array}\right)$} & \multirow{3}{*}{ $3$ } & \multirow{3}{*}{$3$} &  \multirow{3}{*}{$-\frac{1}{3}$
} & 
$u_R^a$                                                                             & $3$       & $1$            & $\frac{2}{3}$ \\ 
&&&& $d_R^a$                                                                             & $3$       & $1$            & $-\frac{1}{3}$ \\ 
&&&& $D_R^{1,\,2}$                                                                       & $3$       & $1$            & $-\frac{4}{3}$  
\\ 
\multirow{3}{*}{$\left(\begin{array}{c}b_L\\ -t_L\\ T_L\end{array}\right)$} & \multirow{3}{*}{ $3$ } & \multirow{3}{*}{$\overline{3}$} &  \multirow{3}{*}{$\frac{2}{3}$
} & 
$T_R$                                                                               & $3$       & $1$            & $ \frac{5}{3}
$\\
&&&& & &    &            \\
&&&&                                                                             &        &             & \\
\multirow{3}{*}{$\left(\begin{array}{c}\ell^a_L\\ -\nu^a_L\\ (\ell^a_R)^c  \end{array}\right)$} & \multirow{3}{*}{$1$} & \multirow{3}{*}{ $\overline{3}$} &  \multirow{3}{*}{$ 0
$} &
$\chi$                                                                              & $1$       & $3$           & $ 1
$ \\
&&&& $\rho$                                                                              & $1$       & $3$           & $ 0
$ \\
&&&& $\eta$                                                                              & $1$       & $3$           & $ -1
$ \\
    \hline
    \hline
\end{tabular}
\caption{\label{tab:fermion} The  fermion and scalar content of the  $A[\sqrt{3}]$ model we are considering. }
\label{table:particlecontent1}
\end{table}

The three BSM fermions $D^{1,2}$ and $T$ are Vector Like Quarks (VLQs).
If we consider the embedding of  $SU_L(2)$ into $SU_L(3)$, we observe that the generators $T^i$ ($i=1,2,3$) do not act on the third component of $SU_L(3)$ triplets and anti-triplets.
As a consequence, 
$D^{1,2}$ and $T$ are singlets of the SM $SU_L(2)$ and they do not couple to charged currents by the covariant derivative.
Later on (see Eq.\,\eqref{eq:lagr_Z}), we explicitly show that their left and right-handed components behave the same way
in the neutral weak interactions, resulting in a vector electroweak current; hence the epithet vector-like.

Being singlets of the SM $SU_L(2)$, VLQs do not require coupling to $SU(2)_L$ scalar Higgs doublets to get mass. Hence they are not excluded by current measurements of the
production rate of the Higgs boson, which instead  exclude the existence of a fourth generation of
chiral quarks.

\subsection{Gauge sector}
\label{Gaugesector}

In the first stage of SSB and with the choice  $\beta=\sqrt{3}$, there emerge singly and doubly charged gauge bosons, $V_\mu^\pm$ (also named $W'_\mu$) and $Y_\mu^{\pm\pm}$. The new charged gauge bosons are dubbed bileptons because they couple at tree level to SM lepton pairs, but not to SM quark pairs~\footnote{To avoid confusion with old literature, we remind that they have previously shared the name dilepton with events having
two final state leptons.}. Bileptons occur in several models of common SM extensions, as grand unified theories, technicolor and compositeness scenarios. We can assign them global quantum numbers, in contrast with the SM where only  fermions carry baryon or lepton numbers, not bosons.

  One distinctive  feature of our 331 model is that  $Y_\mu^{\pm\pm}$ bileptons  couple to BSM quarks. 
  One can assign lepton number $L=-2$ to the gauge bosons $V_\mu^+$ and $Y_\mu^{++}$, and $L=+2$ to $V_\mu^-$ and $Y_\mu^{--}$. The 331 model does not conserve separate family lepton numbers, but the total lepton number is conserved if we assign $L=-2$ to the exotic quarks  $ D^{1,2}$  and $L=+2$ to the $T$ quark.
  Then, 
the global symmetry $B-L$ of the SM maintains  in the minimal model.

After both $SU(3)_L$ breaking eight gauge bosons acquire a mass. 
We assume  that spontaneous symmetry breaking occur   by triplets only, namely the three Higgs fields $\rho$, $\eta$ and $\chi$, defined  in Eq. \eqref{eq:higgstriplet}.
The charged gauge bosons are directly diagonal, while the neutral gauge boson matrix can be diagonalized as shown in appendix B, section \ref{app:Gaugesector}. Thus we have  
\begin{eqnarray}
m_W^2          &   =   & 
\frac{g^2}{4} \left(v^2 + v^{\prime 2}\right)\equiv \frac{g^2}{4} v_+^2\,, \\
m_{V}^2 &   =   & 
\frac{g^2}{4} \left(u^2 + v^{\prime\,2}\right)\simeq \frac{g^2}{4}u^2 \\
m_{Y}^2 &   =   & 
\frac{g^2}{4} \left(u^2 + v^2          \right)\simeq \frac{g^2}{4}u^2 \label{Y_mass}\\
m_A&=& 0\,,\\
m_Z^2          &   \simeq  & 
\frac{g^2}{4\cos^2\theta_W}v_{\rm SM}^2\,\\        
m_{Z^\prime}^2 &   \simeq   & 
\frac{g^2 u^2 \cos^2\theta_W}{3[1-(1+\beta^2)\sin^2\theta_W]}\,.\label{Zprime_mass}
\end{eqnarray}
where  $v$, $v^\prime$ and $u$ are respectively $\rho$, $\eta$, $\chi$ vacuum expectation values, $A_\mu$ is the SM photon field and $\theta_W$ is the SM Weinberg angle (see the appendix for details).
To correctly restore the SM, we require that $v^2_+ = v^2_{\rm SM}$, where $v_{\rm SM}$ is the Higgs boson vacuum expectation value, and $u\gg v, v^{\prime}$. These assumptions imply that $m_W \ll m_V\simeq m_Y$.
Moreover, a useful relation between the masses of these bosons (that one can obtain by comparing Eq.\,\eqref{Zprime_mass} and Eq.\,\eqref{Y_mass} in the appendix) holds:
\begin{equation}\label{eq:ratio-masses}
    \frac{m_{Z^\prime}^2 }{m_Y^2} \simeq \frac{4 \cos^2\theta_W}{3 \left[1 - \left(1+\beta^2\right)\sin^2\theta_W\right] }. 
\end{equation}
Fixing $\beta=\sqrt{3}$ we have 
{\color{magenta}\footnote{The running of the sinus is assumed in the SM framework, see e.g. \cite{Amoroso:2023uux}.}}
\begin{equation}
\label{eq:ratio-masses-1}
    \frac{m_{Z^\prime}^2 }{m_Y^2} \simeq 
    \left\{ 
    \begin{array}{l}
             \,13 \qquad \text{for} \,\sin^2\theta_W(m_Z)\,\,\simeq 0.23  \\
             \,25 \qquad \text{for} \,\sin^2\theta_W(1\;\text{TeV})\simeq 0.24
    \end{array}
 \right.
\end{equation}

\subsection{Yukawa interactions}

SM fermions get mass via the Yukawa coupling with the scalars. The gauge invariant Yukawa Lagrangian (for the $\beta=\sqrt{3}$ case) reads as
\begin{equation}
\begin{split}
\mathcal{L}_{Yukawa} =& +
\lambda_{i,a}^d \overline{Q}_i \rho   d_{a,R} +
\lambda_{i,a}^u \overline{Q}_i \eta   u_{a, R} +
\lambda_{3,a}^d \overline{Q}_3 \eta^* d_{a,R} +
\lambda_{3,a}^u \overline{Q}_3 \rho^* u_{a, R}+\\
&+
\lambda_{i,j}^J \overline{Q}_i \chi  D_{j,R}+\lambda_{3, 3}^J \overline{Q}_3 \chi^* T_{R}
\\
&+
\lambda_{a,b}^\ell \epsilon_{ijk} \overline{L}_{ai} L_{bj} \rho^*_k
\label{eq:yukawa}
\end{split}
\end{equation}
where we have used the notations of the appendix and set $[Q^1,\,Q^2,\,Q^3]\equiv [3_Q^1,\, 3_Q^2,\, \overline{3}_Q^3]$, $[L^1,\,L^2\,,L^3] \equiv [\overline{3}_\ell^1,\, \overline{3}_\ell^2,\, \overline{3}_\ell^3]$, with $a, b = 1, 2, 3$ and $i, j = 1, 2$. The first line corresponds to the masses of the Standard Model quarks, the second to those of the VLQs, and the third pertains to the charged lepton masses. 
The Yukawa interactions involving quarks are 331-invariant independently from the $\beta$, whereas the ones accounting for the charged lepton masses (last line of Eq.\,\eqref{eq:yukawa}) are invariant only for  $\beta=\sqrt{3}$ being $X[\rho] = 0$.

From the diagonalization of Eq.\,\eqref{eq:yukawa}, we obtain the quark mass eigenstates, $u^\prime_{a}$ and $d^\prime_{a}$. 
The relation between left-handed and right-handed mass and interaction eigenstates reads as 
\begin{equation}
\left(\begin{array}{c}u_{1,}^\prime\\u_{2}^\prime\\u_{3}^\prime\end{array}\right)_{L,R} = {\cal U}_{L,R}^{-1}\left(\begin{array}{c}u_{1}\\u_{2}\\u_{3}\end{array}\right)_{L,R}
\hspace{3cm}
\left(\begin{array}{c}d_{1}^\prime\\ d_{2}^\prime\\ d_{3}^\prime\end{array}\right)_{L,R} = {\cal V}_{L,R}^{-1} \left(\begin{array}{c}d_{1}\\d_{2}\\d_{3}\end{array}\right)_{L,R} 
\end{equation}
where ${\cal U}_{L,R}$ and ${\cal V}_{L,R}$ are  four unitary matrices. 
The CKM mixing matrix $V_{\rm CKM}$ is recovered as the usual product $V_{\rm CKM} \equiv {\cal U}_L^\dagger\cdot {\cal V}_L $.

We can safely take  the new quarks $D$ and $T$ as mass eigenstates and omit their rotation matrices in the interactions. That is immediately evident for $T$, which, having charge 5/3, cannot mix with the other quarks. In the case of $D$ quarks, let us observe that  in the interactions a possible rotation  matrix is either cancelled because of unitarity (see Eqs. \eqref{eq:lagr_Z} and \eqref{eq:lagr_Zprime},) or it appears together with  ${\cal U}_L$ or ${\cal V}_L$ (see Eqs. \eqref{eq:lagr_V} and \eqref{eq:lagr_Y}). In the latter case, we can define new unitary ${\cal U}_L$ or ${\cal V}_L$ matrices as  the  product of the older ${\cal U}_L$ or ${\cal V}_L$ matrices, whose entries are unknown, and  the $D$ rotation matrix, similarly unknown. 

\subsection{Fermions and gauge boson new couplings}
This section illustrates new interactions predicted by the model under investigation. We start from the interactions involving the SM gauge bosons and the new fermions. The $W^\pm_\mu$ bosons do not couple to VLQs, while the $Z_\mu$ boson does. The Lagrangian involving the $Z_\mu$ boson reads as 
\begin{equation}
         i \mathcal{L}_F^{Z} \supset + i \frac{g}{c_W} Z^\mu \Bigg[ \frac{4}{3}s_W^2 \overline{D}_L^i\gamma_\mu D_L^i+  \frac{4}{3}s_W^2 \overline{D}_R^i\gamma_\mu D_R^i- \frac{5}{3}s_W^2 \overline{T}_L\gamma_\mu T_L -\frac{5}{3}s_W^2 \overline{T}_R\gamma_\mu T_R\Bigg]
         \label{eq:lagr_Z}
\end{equation}
Here we report the terms involving the $Z^\prime_\mu$ boson
\begin{equation}
\label{eq:lagr_Zprime}
    \begin{split}
        i \mathcal{L}_F^{Z^\prime} =& i \frac{g}{2\sqrt{3} c_W \sqrt{1-4\,s_W^2}} \Biggl\{ \left[1-4\,s_W^2\right] (\overline{\nu}_L^a \gamma_\mu \ \nu_L^a+ \overline{\ell}_L^a \gamma_\mu \ell_L^a)-\left[2+4\, s_W^2\right]\overline{\ell}_R^a \gamma_\mu \ell_R^a +\\
        &+\left[-1 + 2\,s_W^2\right]\overline{u}_L^{\prime\,a} \gamma_\mu \delta_{ab}u_L^{\prime\,b} + 2c_W^2 \overline{u}_L^{\prime\,a} \gamma_\mu u_L^{\prime\,b} {{\cal U}_L}^*_{3a}{{\cal U}_L}_{3b}+4\,s_W^2 \overline{u}_R^{\prime\,a}\gamma_\mu u_R^{\prime\,b}\delta_{ab} +\\
        & + \left[-1 + 2\,s_W^2\right]\overline{d}_L^{\prime\,a} \gamma_\mu \delta_{ab}d_L^{\prime\,b}+ 2c_W^2 \overline{d}_L^{\prime\,a} \gamma_\mu d_L^{\prime\,b} {{\cal V}_L}^*_{3a}{{\cal V}_L}_{3b}- 2 s_W^2 d_R^{\prime\,a} \gamma_\mu d_R^{\prime\,b} \delta_{ab}+ \\
        & + 8\,s_W^2 (\overline{D_L}^i\gamma_\mu D_L^j +\overline{D_R}^i\gamma_\mu D_R^j)+ 10\,s_W^2(\overline{T_L}\gamma_\mu T_L+\overline{T_R}\gamma_\mu T_R) \Biggl\}
    \end{split}
\end{equation}
 While in the SM the GIM mechanism guarantees the flavor universality, $Z^\prime_\mu$ interactions with quarks violate it, allowing processes like $e^+e^- \to \overline{u} c$ or
$\overline{u} u \to \overline{u} c$.

The vertices involving $V_\mu$ read as \begin{equation}
\label{eq:lagr_V}
i \mathcal{L}_F^V = + i \frac{g} {\sqrt{2}} \, V^{-}_\mu \left[\overline{\nu}^a_L \gamma^\mu  (\ell_R^a)^c + \overline{D}_L^i \gamma^\mu d_L^{\prime\,a} {{\cal V}_L}_{ai}+ 
\overline{u}_L^{\prime\,a} \gamma^\mu T_L \, {{\cal U}_L}_{3a}^* +{\rm h.c. }  \right],
\end{equation}
whereas the ones with the $Y_\mu$ boson are 
\begin{equation}
    i \mathcal{L}_F^Y =- i \frac{g}{\sqrt{2}}\, Y^{++}_\mu \left[
    \overline{(\ell_R^a)^c} \gamma^\mu \ell_L^a  
    +\overline{u}_L^a\gamma^\mu D_L^i \, {{\cal U}_L}^*_{ai}
    +\overline{T}_L\gamma^\mu d_L^a \, {{\cal V}_L}_{3a} 
 +{\rm h.c. }  \right]
\label{eq:lagr_Y}
\end{equation}

\noindent
In the appendix, we show the Feynman diagrams of  the couplings which are relevant to our phenomenological analysis.
 
 We observe that the quark couplings to  the new gauge bosons depend on the unknown entries of the rotation matrices ${\cal U}_L$ and ${\cal V}_L$, which do not cancel out in the interactions.
For simplicity in our analysis we fix as benchmark case ${\cal U}_L  = \mathbf{1}$, i.e. the three by three unity matrix; from the definition of the CKM matrix $V_{CKM}$ it follows that ${\cal V}_L = V_{CKM}$. Under this assumption  $D^i$ and $T$ couple predominantly  to the $i$-th quark generation and  to the third, respectively. For our phenomenological analysis, it is not worrisome having set ${\cal U}_L $  real, since a possible phase would  not affect the diagrams in Fig. \ref{fig2}. It is possible to connect the entries of the rotation matrices to observables in kaon, $B_d$, and $B_s$ systems, yielding absolute values that are generally of the same order of magnitude than ours
\cite{Buras:2012dp}.

\section{Phenomenology of minimal 331}\label{sec:pheno}

This minimal 331 model  is rich of new fields. 
In the fermion sector we have two VLQs and in the gauge sector we have one new neutral  and two new charged bosons, namely
\begin{equation}
    (D^{1,\,2})^{-4/3},\,T^{5/3},\,Z^{\prime\, 0},\,V^{\pm},\,Y^{\pm\pm},
\end{equation}
where we added an apex to indicate the charge of the particle. 

Several phenomenological implications 
of the existence of these exotic particles 
have been investigated: 
\begin{itemize}
\item At colliders, VLQs are typically identified 
by means of the decay channels
    $\rm{VLQs}\to q\,W^{\pm},\,q\,Z,\,q\,H$ \cite{ATLAS:2015ktd, CMS:2017ynm, ATLAS:2018tnt, ATLAS:2022hnn,ATLAS:2023pja}.
In 331 models these decay channels are suppressed and are replaced by the dominant ones $\rm{VLQs}\to q\,V^{\pm},\,q\,Y^{\pm\pm}$.
Therefore 
 the peculiar interactions of VLQs in 331 theories allow less  conventional search strategies, 
see for instance Ref. \cite{Corcella:2021mdl}, where the case of an exotic VLQ with charge 5/3 has been considered and a lower bound on its mass of about 1.5 TeV is found by a recasting of the CMS analysis in Ref. \cite{CMS:2018ubm}. 

\item 
The phenomenology of $Z^\prime$ bosons in 331 has been studied  quite in detail in the recent literature \cite{Oliveira:2021gcw, Alves:2022hcp,Addazi:2022frt, Alves:2023vig}. 
The $Z^\prime$ boson is produced through the Drell-Yan process and subsequently decays into pairs of SM particles or in exotic channels, like in pairs of VLQs. Note that these signatures are typical also of  $Z^\prime$ bosons or, in general, of heavy resonances, which appear in other BSM theories, as those of composite Higgs models \cite{Contino:2006nn,Vignaroli:2014bpa, Vignaroli:2015ama, Bini:2011zb}. Therefore, it would be beneficial to identify, as we will do in this study, different signatures which are more characteristic of the 331 model. The dilepton channel gives significant constraints on the new vector, with bounds of the order of 4 TeV on the $Z^\prime$ mass, 
as extracted 
from the ATLAS and CMS searches, already with a luminosity of 139 fb$^{-1}$ \cite{Alves:2022hcp}. Note, however, that the $Z^\prime$ mass is expected to be significantly higher than those of $V$ and $Y$ (Eq. \ref{eq:ratio-masses-1}). While it will certainly be useful to continue in these searches for the $Z^\prime$, it is essential to explore new strategies for testing the 331 model that can be more distinctive of the underlying theory and, possibly, even more efficient.

\item 

Differently from the typical LHC phenomenology of $W^\prime$ resonances (such as Sequential-SM $W^\prime$ \cite{Altarelli:1989ff}, Kaluza-Klein $W$ from extra-dimension or from composite Higgs theories \cite{Contino:2006nn}), which are dominantly produced via the Drell-Yan process \footnote{The production via weak-boson-fusion is typically subdominant, except when considering colliders at very high collision energies, such as FCC-hh \cite{Mohan:2015doa}}, in the case of the $V^{\pm}$ ($W^\prime$) from 331 models, other production channels must be considered.
 Indeed, 
their Drell-Yan production is precluded at tree level, given the absence of $V$ couplings with pairs of SM fermions, see e.g. Ref. \cite{Corcella:2021mdl}. Nevertheless, the $V^{\pm}$ could be produced by interactions with a VLQ and a SM quark.
It could be thus interesting to examine less conventional channels, such as the one in Fig.\,\ref{fig0} (bottom) in which the $V$ is produced in association with a VLQ. We leave this analysis to a future investigation.

\end{itemize}

In our study we
propose signatures for 331 models which  
are also efficient as search channels for the exotic VLQs.
In particular, based on the considerations above,
we find particularly interesting to focus on the {\it inclusive} production of a pair of bilepton vector bosons $Y^{\pm\pm}$ 
\begin{equation}\label{process}
pp\to Y^{++}Y^{--} 
\end{equation}
\noindent where each bilepton  decays  into a pair of same sign charged leptons. Then the process we look for is 
\begin{equation}
    pp\to \ell^+\ell^+\,\ell^-\ell^- +{ X}
\end{equation} 
with a multi-lepton final state. We consider the {\it inclusive} production, allowing the presence of extra jets in the final state denoted here as $X$.

\begin{figure}[ht!]
\centering
\includegraphics[scale=0.9]{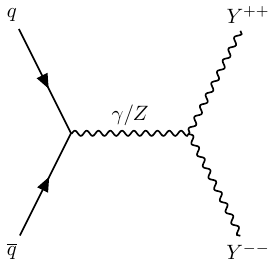}
\qquad \qquad
\includegraphics[scale=0.9]{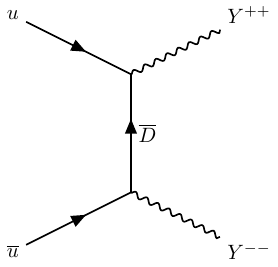} \caption{Typical Feynman diagrams for $Y^{++}Y^{--}$ production: Drell-Yan like processes (left), t-channel mediated by  a vector-like anti-quark (right).}\label{fig1}
\end{figure}

\begin{figure}[ht!]
\centering
\includegraphics[width=0.3\textwidth]{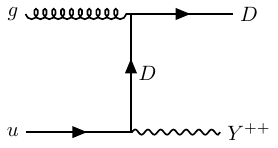} \qquad \qquad\includegraphics[width=0.3\textwidth]{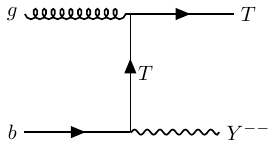} \\
\includegraphics[width=0.3\textwidth]{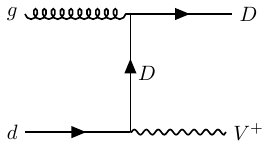} \qquad \qquad\includegraphics[width=0.3\textwidth]{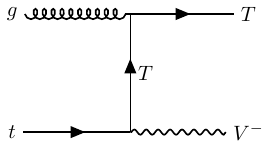} 
\caption{ Bilepton $Y$ and $V$ associated production with a VLQ.
\label{fig0}
} 
\end{figure}

Exclusive production of bilepton pairs  at colliders, with purely $Y^{++}Y^{--}$ in the final state, has been already studied, 
see for instance Refs.
\cite{Dion:1998pw, Corcella:2017dns, Corcella:2018eib, Nepomuceno:2019eaz}. 
The typical diagram considered (at the parton level) are the Drell-Yan like processes (Figure\,\ref{fig1}, left) or with the exchange of an exotic quark in the $t$-channel (Figure\,\ref{fig1}, right). 

Another 
possibility \cite{Dutta:1994pd} (not considered in the previous searches for exclusive bilepton pair production)
is  the production from gluon ($g$) interaction, associated to a VLQ, namely  $g\,u\to D^1 Y^{++}$, $g\,c\to D^2 Y^{++}$ and $g\,b\to T Y^{--}$, as shown in Figure \ref{fig0}. 
If
\begin{equation}\label{eq:hierarchy1}
  m_Y< m_{D},\,m_T  
\end{equation}
then the VLQs can decay in bileptons as $D^1\to Y^{--}u$, $D^2\to Y^{--}c$ and $T\to Y^{++}b$, see Figure \ref{fig2}. The decay widths are reported in the appendix.

\begin{figure}[h!]
\centering
\includegraphics[scale=0.9]{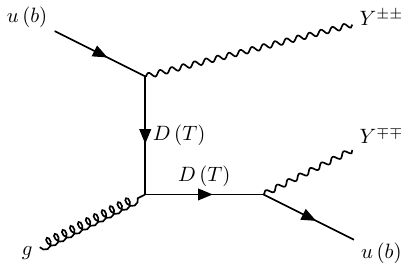} 
\includegraphics[scale=0.9]{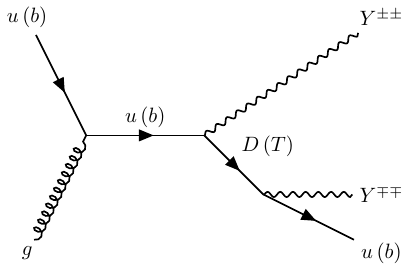} \\
\includegraphics[scale=0.9]{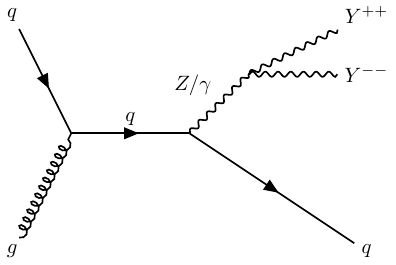} 
\includegraphics[scale=0.9]{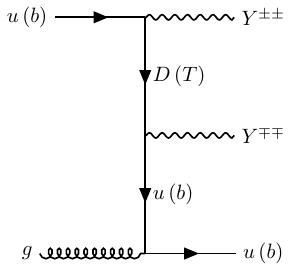}\\ 
\includegraphics[scale=0.9]{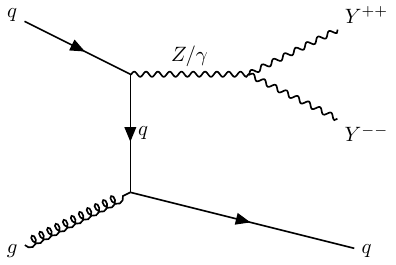} 
\caption{Feynman diagrams for the dominant contributions to the process $g\,q \to Y^{++}Y^{--}\,j$ given by the VLQ-Y associated production considered in this study. The symbols $g$, and $q$ label a  gluon and a SM quark, respectively; $j=q,\,g$ at the parton level. }\label{fig2}
\end{figure}

A contribution to the inclusive process \eqref{process} is therefore also given by 
\begin{equation}\label{eq:unjet}
    g\,q \to {\rm VLQ} \; Y \to Y^{++}Y^{--}\, q^\prime,
\end{equation}
where the VLQ is either a  $D$ or a $T$ fermion; $q$ and $q^\prime$ are either an up, a charm or a bottom SM quark. 
We remark that these processes are not included among the ones considered in Ref.
\cite{Corcella:2017dns}, which focuses on the exclusive final state 
$Y^{++}Y^{--}\,jj$ with $j=q,\,g$ at the parton level.  Conservatively, in our study we will not include the $D^2$ contribution (i.e. we will consider either $D^2$ decoupled from the spectrum or $m_Y> m_{D^2}$ so that the $D^2 \to Y c$ decay is off-shell and suppressed). This is a conservative choice, since an even larger VLQ contribution to the bilepton pair cross section is expected when $D^2$ can contribute. In agreement with  this assumption, we indicate for simplicity $D^1\equiv D$ hereafter and in the  Fig.  \ref{fig1} and \ref{fig0}.

\begin{figure}[ht!]
\centering
\includegraphics[width=0.44\textwidth]{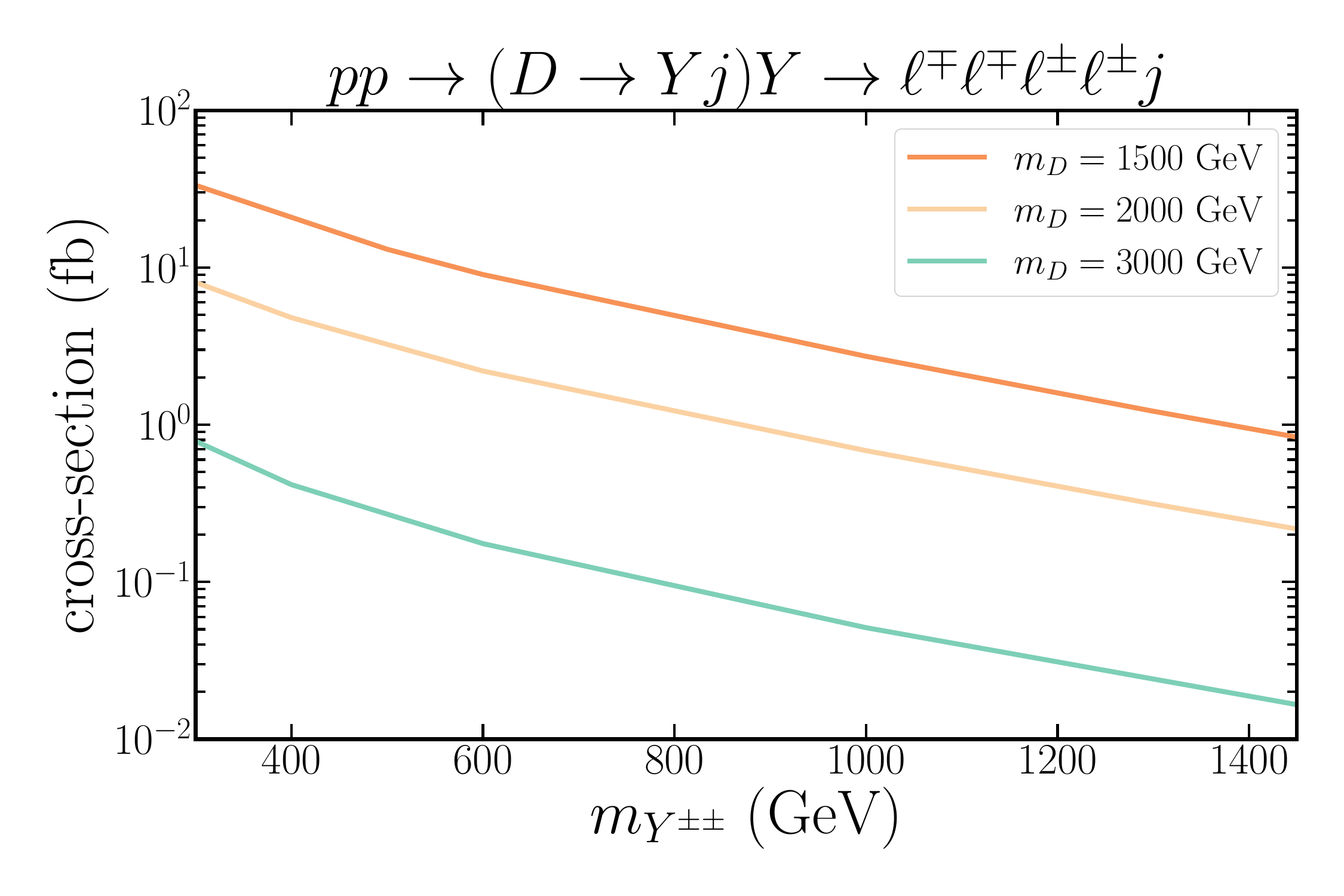} \; \includegraphics[width=0.45\textwidth]{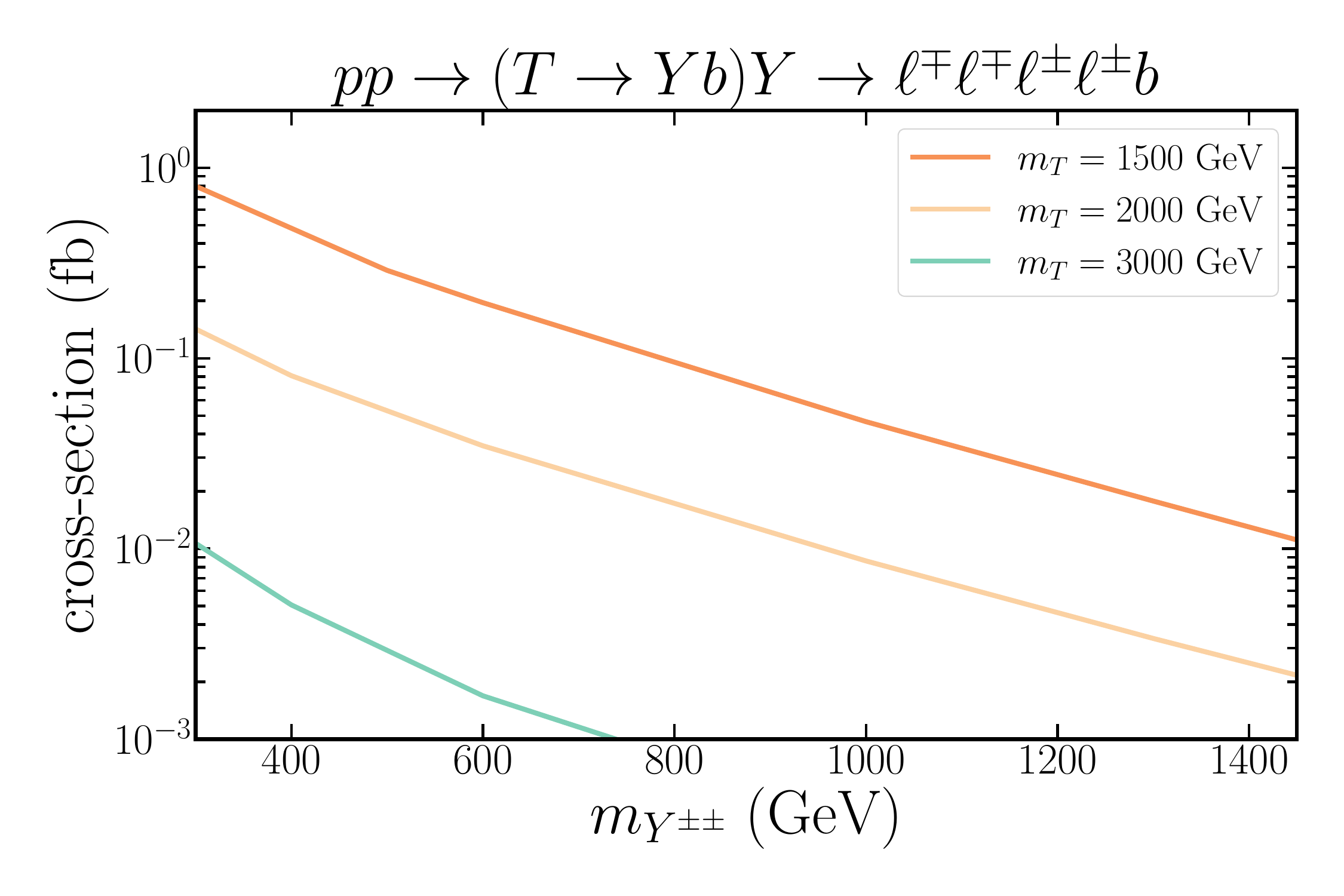} 
\caption{Cross section for the associated production of a bilepton with a VLQ. We consider the subsequent decay of the VLQ into a bilepton plus a quark, and the bilepton decay into electron or muon pairs. \label{fig:xsec-VLQ-Y}}
\end{figure}

\begin{figure}[]
\centering
\includegraphics[width=0.6\textwidth]{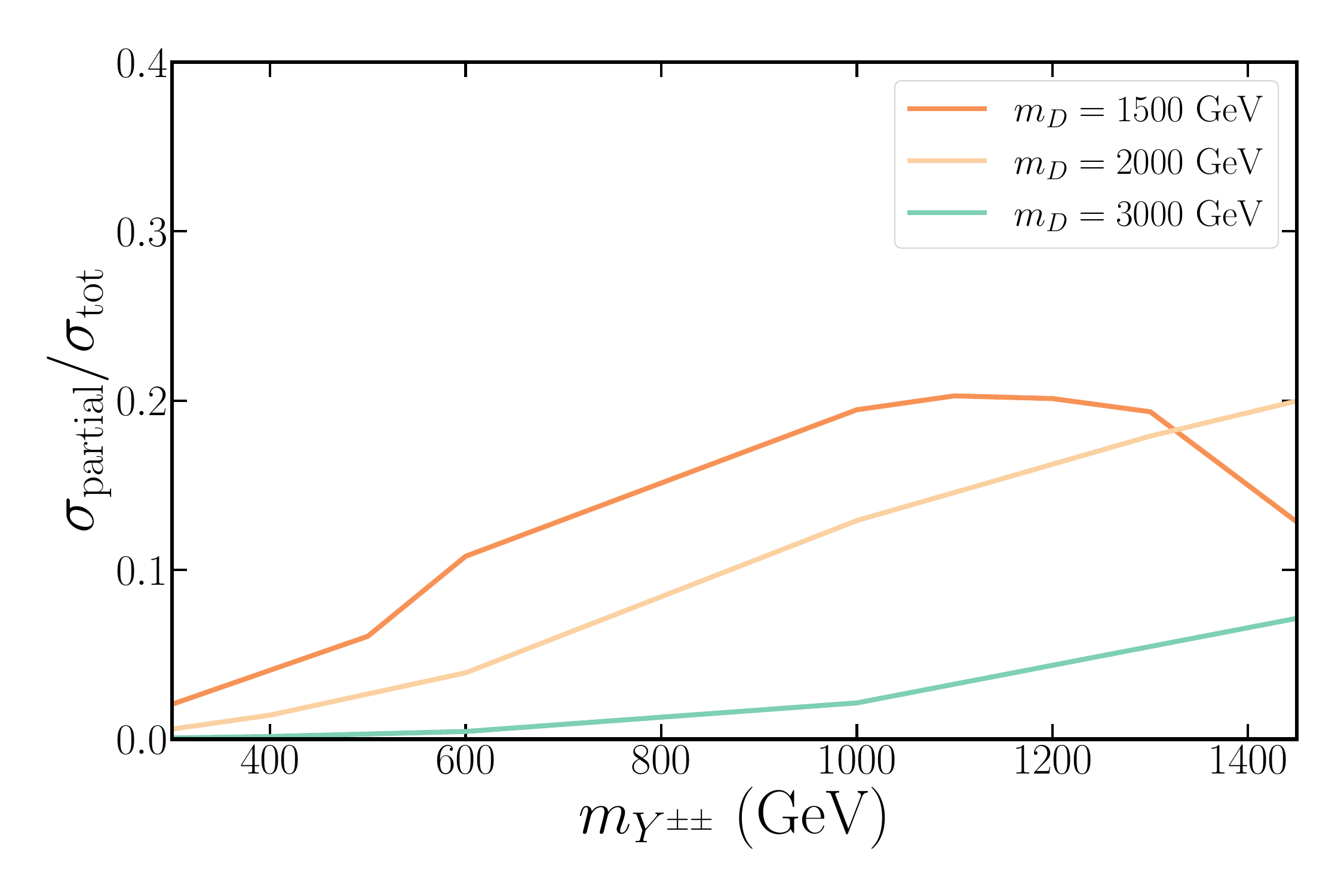} \; 
\caption{\label{fig:associated-xsec} Ratio of the partial cross section for  the bilepton associated production with a $D$ VLQ, which subsequently decays into a second bilepton, i.e. the $D$-$Y$ $\to YYj$ associated production ($\sigma_{\rm partial}$) over the total $YY+X$ cross section ($\sigma_{\rm tot}$). }
\end{figure}

To our knowledge, all previous studies have overlooked the contribution to the pair bilepton production given by the bilepton associated production with a  VLQ (Fig. \ref{fig0}), which subsequently decays into a second bilepton, see Fig. \ref{fig2}. 
We find that instead this contribution is significant in a large portion of the model parameter space. We show in Fig. \ref{fig:xsec-VLQ-Y} the cross section for the VLQ-bilepton associated production, for the cases of $D$ and $T$ VLQ, as a function of the bilepton mass and for different VLQ mass values. Because of the smaller PDF for the initial b-quark in the $T-Y$ associated production process, the latter has a significantly lower cross section compared to the $D-Y$ associated production. An analogous consideration applies to the case of the $D^2-Y$ associated production, which depends on the $c$-quark PDF. Nevertheless, identifying these specific associated productions, possibly by applying a tailored analysis at a high luminosity, could give important confirmation about the flavor structure of the 331 model. In the case of the $D$ VLQ, the associated production with a bilepton constitutes up to $\sim$20\% of the total inclusive cross section for the bilepton pair production in a large portion of the parameter space, in particular when $m_Y$ approaches $m_D$ (cfr. Fig. \ref{fig:associated-xsec}). On top of their contribution to the inclusive bilepton pair production, these channels could be considered for a specific exclusive analysis. In this case, one could select the signal from the background by exploiting the peculiarities of the VLQ-Y associated production, namely the presence in the final state of a heavy VLQ (whose invariant mass could be reconstructed), a jet and bileptons with hard $p_T$ and peculiar angular distributions. We leave this analysis for a future project.

\begin{figure}[h]
\centering
\includegraphics[width=0.4\textwidth]{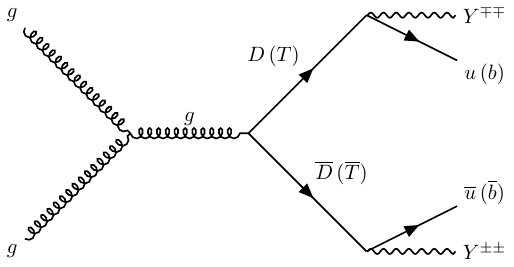} 
\caption{ \small 
\label{fig:pair_prod} QCD pair production of VLQs, leading to a $YYjj$ final state.
} 
\end{figure}

Finally, another contribution to the inclusive bilepton pair production, which we have included in our study, come from the QCD pair production of extra VLQs, which both decay into a bilepton plus a jet: $pp \to ({\rm VLQ} \to Y \, q) \; ({\rm VLQ} \to Y q) \to YY jj $, with VLQ=$D,T$ and $j=u,b$, see Figure\,\ref{fig:pair_prod}.  

\section{Bilepton pair production at LHC}\label{sec:bilepton_prod}

Given the phenomenological considerations expressed in the previous section, we find particularly promising the search for bileptons at the LHC.
We want to focus in particular on the inclusive process $pp\to \ell^{\pm}\ell^{\pm} \ell^{\mp}\ell^{\mp}$,  where a pair of $Y$ is produced on-shell and then decay to same sign leptons, allowing the presence of extra jets in the final state. 
We observe that ${\rm Br}(Y^{\pm\pm}\to \ell^{\pm}\ell^{\pm}) \simeq 1$ because this is the only decay channel at tree level with our mass choice.  
Our analysis consistently considers all possible contributions to the inclusive bilepton pair production, and includes the processes whose diagrams are shown in Figures \ref{fig1}, \ref{fig2}, \ref{fig:pair_prod}. 
For our analysis, we have implemented, by means of {\textsf Feynrules} \cite{Alloul:2013bka}, the 331 model we are considering
($A[\sqrt{3}]$) 
in a {\textsf Universal Feynrules Output} model \cite{Degrande:2011ua}, which we use for simulations and numeric calculations in {\textsf MadGraph5} \cite{Alwall:2014hca}.

\begin{figure}[ht!]
\centering
\includegraphics[width=0.65\textwidth]{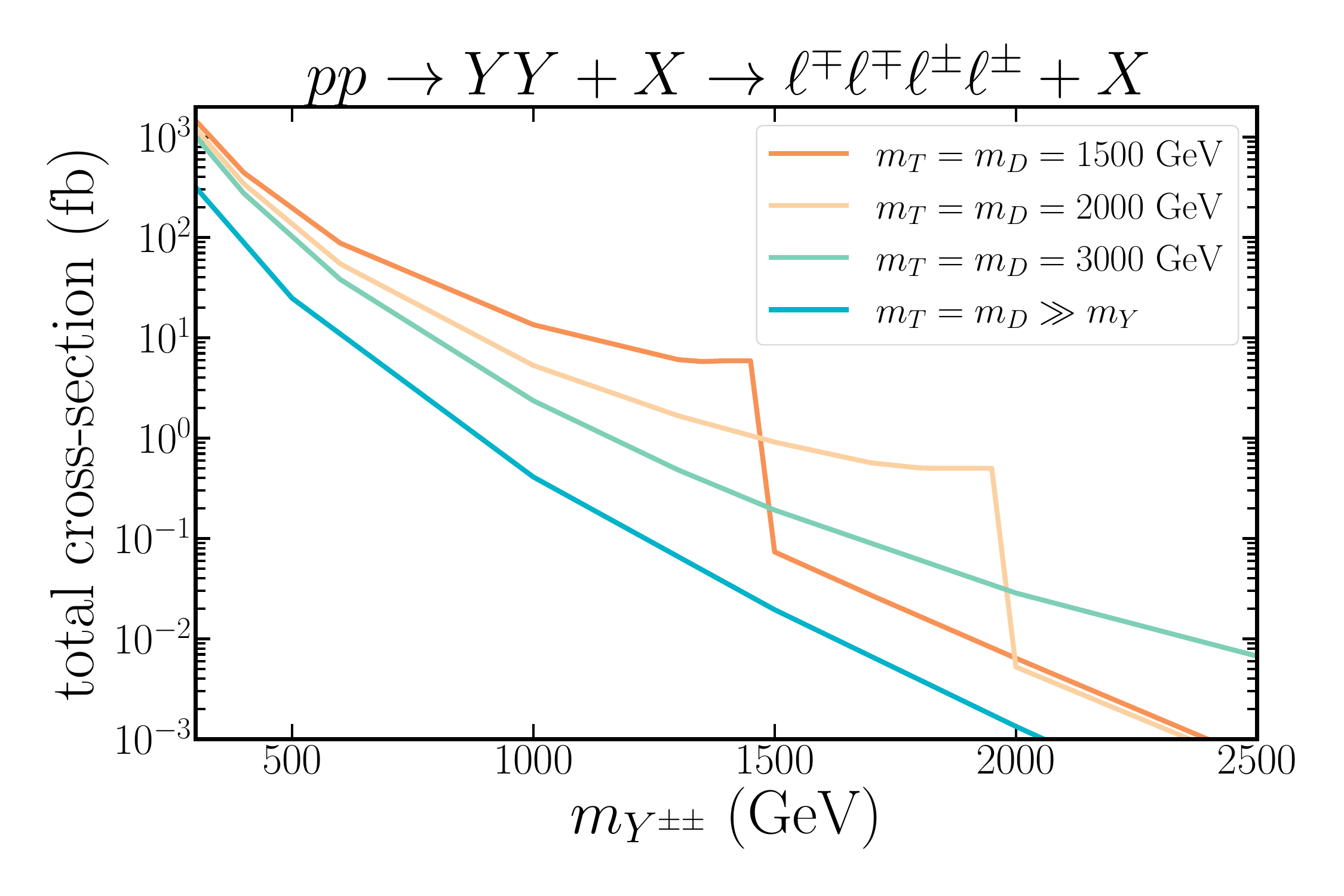} 
\caption{ Inclusive cross section for the process $pp\to \ell^{\pm}\ell^{\pm} \ell^{\mp}\ell^{\mp}+X$
as function of the Y mass and for different values of the VLQ masses, and in the case
of decoupled VLQ.\label{fig:xsec-inclusive}}
\end{figure}
In Figure\,\ref{fig:xsec-inclusive} we provide the inclusive cross section for the process
$pp\to \ell^{\pm}\ell^{\pm} \ell^{\mp}\ell^{\mp}+X$ as function of the $Y$ mass and for different values of the VLQs masses, and in the case of decoupled VLQs.
It is evident that processes involving VLQs which decay into bileptons constitute the dominant contribution to the bilepton pair inclusive cross section. Indeed, a large boost in the cross section, which reaches up to about two orders of magnitude near the kinematic threshold $m_{VLQ}\simeq m_Y$, is observed when the $VLQ\to Yq$ decays are on-shell. When these decays become off-shell, the cross section reduces significantly, despite a residual contribution from VLQ exchanged in the t-channel is still relevant. 

A recent ATLAS search for doubly charged Higgs bosons, at the 13 TeV LHC with 139 fb$^{-1}$, focus on multilepton final states with at least one same-sign lepton pair \cite{ATLAS:2022pbd}. This analysis is therefore sensitive also to the bilepton pair production.
We apply a recast of this analysis to extract a lower limit on the $Y$ mass.

\subsection{Reinterpretation of the ATLAS search for doubly charged Higgs bosons}

We summarize in the following paragraphs the main selection criteria considered in the ATLAS analysis \cite{ATLAS:2022pbd}, also adopting their notation. The search  considers a final state with two (SR2L), three (SR3L) or four (SR4L) leptons, either electrons or muons, passing triggering requirements. The leptons must satisfy the transverse momentum $p_T$ and rapidity $\eta$ requirements: $p_T>$ 40 GeV and $|\eta_\mu|<$2.5, $|\eta_e|<$2.47 (for the electron, the region 1.37$<|\eta_e|<$1.52 is vetoed). Electron candidates are discarded if their angular distance $\Delta R=\sqrt{\Delta \phi^2 +\Delta \eta^2 }$  from a jet (with $p_T>$20 GeV and $|\eta|<$2.5) satisfies 0.2$<\Delta R<$0.4. If a muon and a jet featuring more than three tracks have a separation $\Delta R<$0.4 and the muon's $p_T$ is less than half of the jet’s $p_T$, the muon is discarded.
Signal regions are optimised to search for one same-charge lepton pair, in the case of two or three leptons, while four-lepton events are required to feature two same-charge pairs. The signal selection exploits, as the main variable to distinguish signal events from the background, the invariant mass of the two same-charge leptons with the highest $p_T$ in the event, $m(\ell^{\pm},\ell^{\prime\pm})_{\rm lead}$, and applies a cut:
\begin{equation}
m(\ell^{\pm},\ell^{\prime\pm})_{\rm lead} \geq 300 \; {\rm GeV} \;.
\end{equation}
Furthermore, for SR2L and SR3L, the vector sum of the two leading leptons’ transverse momenta must exceed 300 GeV:
\begin{equation}
p_T(\ell^{\pm},\ell^{\prime\pm})_{\rm lead} \geq 300 \; {\rm GeV} \; \; {\rm (SR2L, SR3L)}
\end{equation}
while, for SR4L, the average invariant mass of the two same-charge lepton pairs must satisfy:
\begin{equation}
\bar{m}=(m_{\ell^{+}\ell^{\prime +}} +m_{\ell^{-}\ell^{\prime -}} )/2 \geq 300 \; {\rm GeV} \;\;  {\rm (SR4L)}.
\end{equation}
On top of these cuts, for SR3L and SR4L, a $Z$-veto condition is required, excluding the presence of same-charge lepton pairs with invariant mass 71 GeV$< m(\ell,\ell^\prime)<$111.2 GeV. For SR2L, the requirement on the separation $\Delta R(\ell^{\pm},\ell^{\prime\pm})_{\rm lead}<3.5$ is considered.
Finally,  events containing jets identified as originating from b-quarks are vetoed.

We apply the above selection to the signal of inclusive bilepton pair production. Events are generated with {\textsf Madgraph5} and then passed to {\textsf Pythia8} \cite{Sjostrand:2014zea} for showering and hadronization. Jets are clustered by using {\textsf Fastjet} \cite{Cacciari:2011ma} with an anti-kt algorithm with radius parameter $R=0.4$. Conservatively, based on the details of lepton reconstruction procedures described in \cite{ATLAS:2022pbd}, we also consider the following identification efficiencies for an electron: 0.58, and for a muon: 0.95. 

We obtain the overall signal efficiencies reported in Table \ref{tab-eff} for several signal benchmark points (we also list the values for the separate efficiencies in the SR4L and SR2L regions). The b-veto applied in the ATLAS analysis completely suppresses the contribution from the VLQ T to the bilepton pair production.  

\begin{table}
\begin{tabular}{|c|c|c|c|c|}
\hline 
$m_Y$ (GeV) & 300 & 600 & 1000 & 1300 \\
\hline
$m_D$=1.5 TeV & 0.14 $^{0.16}_{0.11}$ & 0.37 $^{0.37}_{0.11}$ & 0.36 $^{0.36}_{0.07}$ & 0.33 $^{0.32}_{0.11}$ \\
\hline
$m_D$=2.0 TeV & 0.14 $^{0.13}_{0.10}$ & 0.38 $^{0.37}_{0.08}$& 0.38 $^{0.37}_{0.07}$ &  0.37 $^{0.36}_{0.14}$\\
\hline
$m_D$=3.0 TeV & 0.13 $^{0.13}_{0.07}$ & 0.38 $^{0.38}_{0.10}$ & 0.38 $^{0.37}_{0.09}$ & 0.39 $^{0.38}_{0.08}$ \\
\hline
$m_D \gg m_Y$ &  0.13 $^{0.13}_{0.06}$ & 0.38 $^{0.38}_{0.09}$ & 0.38 $^{0.37}_{0.09}$ & 0.39 $^{0.38}_{0.08}$ \\
\hline
\end{tabular}
\caption{Efficiencies of the inclusive bilepton pair production to the ATLAS selection \cite{ATLAS:2022pbd} for several benchmark points of the minimal 331 model ($A[\sqrt{3}]$). We report the combined efficiencies for all of the three signal regions considered in the analysis, together with the specific efficiencies for the signal regions SR4L (upper values) and SR2L (lower values). The efficiencies for SR3L are similar to those obtained for SR4L.} \label{tab-eff}
\end{table}

\begin{figure}[ht!]
\centering
\includegraphics[width=0.7\textwidth]{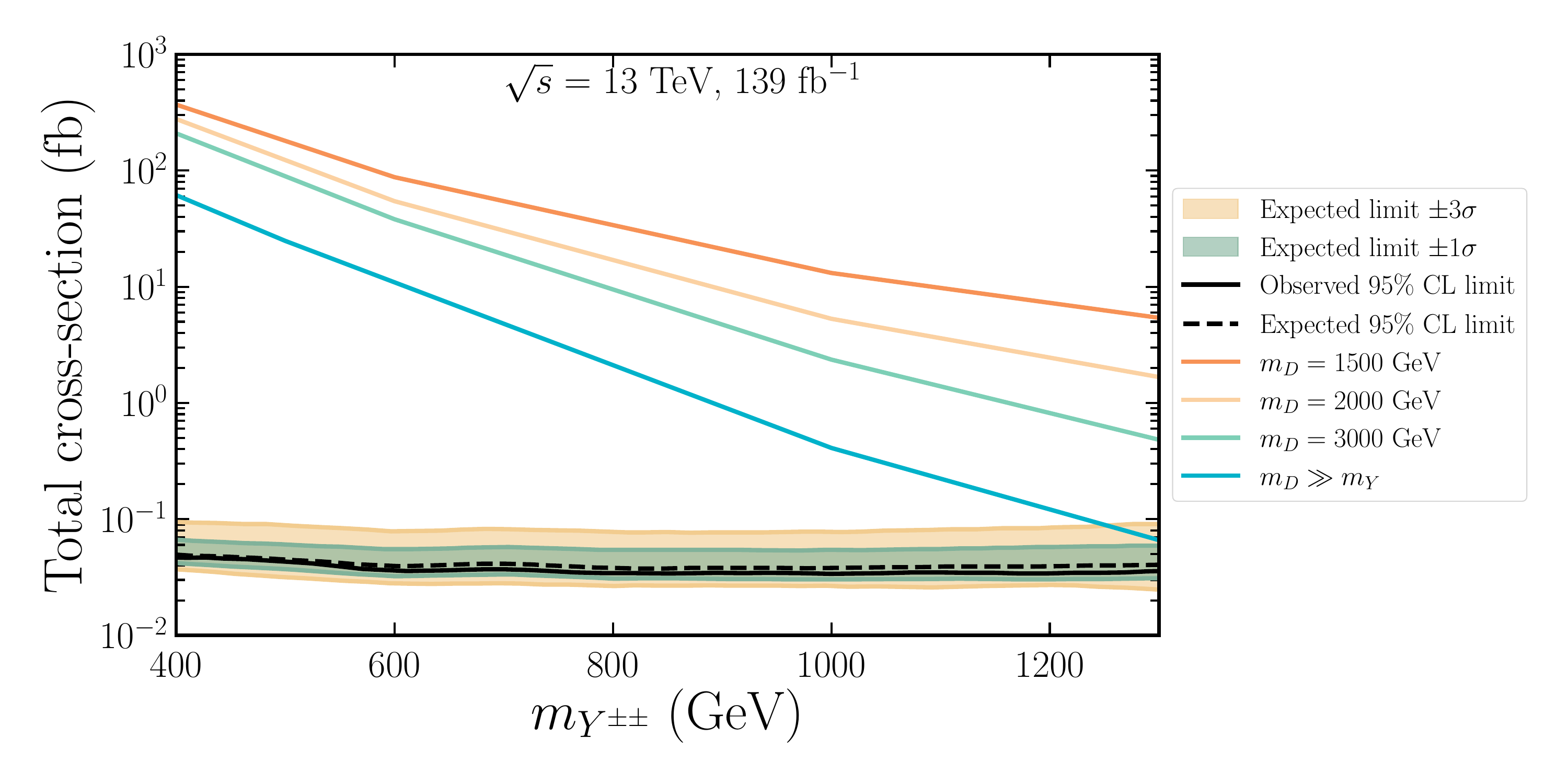} 
\caption{Alongside our prediction for the multi-lepton cross section generated by the bilepton pair production in the minimal 331 model, we illustrate the results obtained by ATLAS in Ref.\,\cite{ATLAS:2022pbd}. In particular, the solid and dashed lines represent the observed and expected 90\% CL upper limits on the multi-lepton cross section as a function of the bilepton mass (doubly charged Higgs in the ATLAS analysis). The shaded green and yellow bands represent the $1\sigma$ and $3\sigma$ uncertainty around the combined expected limit. \label{fig:recast}}
\end{figure}

For all of the benchmark points analyzed, except the lowest mass point, we find efficiencies for the bilepton pair process ($\epsilon_{YY}$) which are comparable to those for doubly-charged Higgs bosons ($\epsilon_{HH}$) in the SR3L and in the SR4L regions. For the SR2L region, we find in general slightly lower efficiencies than those obtained for the charged Higgs signal. In order to correctly compare the cross section for our process $pp \to YY  \to \ell^{\pm}\ell^{\pm} \ell^{\mp}\ell^{\mp}$ with the upper limit band on the multi-lepton cross section reported in Fig. 8 of the ATLAS search \cite{ATLAS:2022pbd}, we consider a scaling of our cross section values by a factor $\epsilon_{YY}/\epsilon_{HH}$. Conservatively, the scaling is not applied when this factor is greater than 1. We show the result of the comparison in Fig. \ref{fig:recast}. We find that in the full mass region covered by the ATLAS analysis, and for all of the considered values of the mass of extra VLQs, the cross section for the process of inclusive bilepton pair production lies above the curve representing the 90\% C.L. upper limit, with the lowest values obtained for the scenario of decoupled VLQs. We can therefore extract, independently of VLQ masses, the 90\% C.L. bound on the bilepton mass:

\begin{equation}
m_Y > 1300 \; {\rm GeV}
\end{equation}

This value supersedes the one obtained in \cite{Nepomuceno:2019eaz}, $m_Y>1200$ GeV, which was based on the results of a previous ATLAS search for doubly charged Higgs bosons in multi-lepton final states with 36.1 fb$^{-1}$ at the 13 TeV LHC \cite{ATLAS:2017xqs}.
It is clear from Fig. \ref{fig:recast} that the search, with the increasing of the luminosity expected at the High-Luminosity LHC phase, has a great potential to test even larger bilepton (and VLQs) mass values, especially when the intertwined contribution of VLQs  is taken into account.

\section{Conclusions}
\label{sec:conclusions}

We  have considered a class of models, known as 331 models, characterized by the electroweak gauge group   $SU_L(3) \times U(1)$. We have selected a  minimal 
331 model satisfying  four general requests: 
1) all matter multiplets include SM fermions, namely there are not  multiplets including only BSM fermions; 2) all gauge bosons  have integer charges; 3) all right-handed fermions are $SU(3)_L$ singlet and possess a left-handed counterpart 4) no gauge anomalies. These assumptions constrain   the number of families: one possibility, although not the only one, is that it is  a multiple of the color number.
The model we have chosen  has $\beta=\sqrt{3}$,
 the first two generations
of quarks 
that transform as triplets under $SU_L(3)$, and the third generation of quark, together with the three lepton generations, that transform as anti-triplets. 
In addition to the usual SM fields, this model accommodates some BSM fields, namely  three VLQs, $D^{1,2}$ and $T$, one neutral  gauge boson $Z^\prime_\mu$, the charged boson $V^\pm_\mu$, and the doubly-charged bilepton $Y^{\pm \pm}_\mu$.
The intertwined phenomenology of bileptons and VLQs  looks particularly promising for the search and characterization of 331 models at the LHC.
In our analysis, we propose bilepton pair signatures for 331 models, which are also efficient as search
channels for  VLQs. In particular, 
we focus on the {\it inclusive} production of a pair of doubly charged bileptons,  
 followed by the decay of each bilepton    into a pair of same-sign charged leptons, that is the process $pp\to Y^{++}Y^{--} \to \ell^+\ell^+\,\ell^-\ell^- +{\rm X}$.
The final state of this inclusive process  has multi-leptons (with at least one same-sign lepton pair), and possibly  extra jets.

 In our analysis, we provide the values of the inclusive cross section for the process $pp\to  \ell^+\ell^+\,\ell^-\ell^- +{\rm X}$, as function of the doubly charged bilepton mass.
We include consistently the production of doubly charged bileptons from gluon interaction, involving VLQs.
A highly massive VLQ can decay into a bilepton  and a SM quark $q$, yielding a final state $Y^{++}Y^{--} q$ or $Y^{++}Y^{--} qq$, which contribute to the inclusive process under investigation.
At the best of our knowledge, the production from gluon-quark interaction of a bilepton associated to a VLQ 
has not been  taken into account in  previous searches for inclusive or   exclusive bilepton pair production.
 We find that instead this contribution is significant in a large portion of the model parameter space (up to $\sim20\%$ of the total inclusive cross section for the bilepton pair production).

A recent ATLAS search for doubly-charged Higgs bosons focuses on multi-lepton final states with at least one same-sign lepton pair \cite{ATLAS:2022pbd}. We  apply a recast of this analysis to extract a lower limit on the $Y$ mass. We found, independently of the VLQs masses, a 90\% CL lower limit on the mass of the doubly charged bilepton:
\begin{equation}
    m_Y> 1.3 \, \, \mathrm{TeV}.
\end{equation}
This value supersedes the previous one of $m_Y>1.2$ TeV \cite{Nepomuceno:2019eaz},  based on 2017 ATLAS results.

We conclude by  remarking that the  channel we have investigated remains relevant with the increase of luminosity expected during the High-Luminosity LHC phase, due to its potential to test quite large bilepton (and VLQs) mass values.

\section*{Acknowledgements}
SM and GR acknowledge the support of the research projects TAsP (Theoretical Astroparticle Physics) and ENP (Exploring New Physics), funded by INFN (Istituto Nazionale di Fisica Nucleare), respectively. The work of NV is supported by ICSC – Centro Nazionale di Ricerca in High Performance Computing, Big Data and Quantum Computing, funded by 
European Union – NextGenerationEU, reference code CN\_00000013.

\bibliographystyle{apsrev4-1}
\bibliography{Biblio}

\newpage

\appendix

\section{Basic features of 331 models}\label{sec:Basic_features}

In this appendix, we present in more detail the general features of the 331 models. 
We are using the same convention of  Ref.\,\cite{Buras:2012dp}.
We do not discuss $SU(3)_C$  which does not  change with respect to the SM.
The generators of the $SU(3)_\text{L}$ gauge group are  indicated with $\hat T^1, \cdots, \hat T^8$. The generators of the fundamental representation of $SU(3)_L$ are $ T^a=\lambda^a/ 2$, where 
${\lambda}^i$  are the eight Gell-Mann matrices.  For the $\bar 3$ representation, the generators are  $-( T^a)^T$.
The normalisation of the generators is $\mathrm{Tr}[ T^i \,  T^j]=\delta^{ij}/2 $,  and ${\mathds 1} = \mathrm{diag} (1, 1, 1)$  is the identity matrix in the fundamental space.
It is common to define the $U(1)_X$ generator  as $ T^9 = {\mathds 1}/\sqrt{6}$. With this normalization $\mathrm{Tr}[ T^9 \,  T^9]=1/2 $, similarly to the one of $SU(3)_\text{L}$ generators.\\
In analogy with the SM the electric charge is defined as the diagonal operator
\begin{equation}
    \hat{\mathcal{Q}}=  \hat T^3 +  \frac{\hat{Y}}{2}
\end{equation}
where we  identify the hypercharge operator ${\hat Y}$ as in Eq. \eqref{hyper:ID}
\begin{equation}
	\frac{\hat Y}{2} =\beta \hat T^8+X{\mathds 1}
\label{hyper:ID2}
\end{equation}\\
In the SM left handed fermions are assigned to doublets of $SU_L(2)$. Similarly in 331 models one assigns left handed fields to fundamental irreducible representations ($3$ or $\overline{3}$) of $SU_L(3)$. We indicate them generically with $3_{Q,\ell}$ and $\overline{3}_{Q,\ell}$, distinguishing between quark triplets and anti-triplets ($3_{Q}$ and $\overline{3}_{Q}$) and lepton  ones ($3_{\ell}$ and $\overline{3}_{\ell}$). Thus, for  a generic  SM quark-doublet $(u_L, d_L)$ we set
\begin{equation}\label{assignment}
3_Q =\begin{pmatrix} 
u_L \\ d_L \\ \psi_{L}
\end{pmatrix}_i, 
\qquad
\overline{3}_Q =\begin{pmatrix} 
d_L \\ -u_L \\ \Psi_{L}
\end{pmatrix}_j.
\end{equation}\\

In a similar way for  a generic  SM lepton-doublet $(\nu_L, \ell_L)$ we set
\begin{equation}\label{assignment2}
3_\ell =
\begin{pmatrix} 
\nu_L \\ \ell_L \\  \xi_{L}
\end{pmatrix}_k,
\qquad
\overline{3}_\ell =
\begin{pmatrix} 
\ell_L \\ -\nu_L \\ \zeta_{L}
\end{pmatrix}_r.
\end{equation}
The field assigned to the third components of these triplets are the 
left handed component of a field beyond the SM, namely we have introduced new fields $\psi_i, \Psi_j, \xi_k, \zeta_r$. The indexes $i,j,k,r$ refer to flavor. \\

The action of the charge operator $\hat{\mathcal{Q}}$ on 
$3$ and  $\bar 3 $  representations is given by the eigenvalues of the eigenstates $|3_{Q,\ell}\rangle$, and $|\overline{3}_{Q,\ell}\rangle$ that we indicate, respectively, as $\mathcal{Q}[3_{Q,\ell}]$ and $\mathcal{Q}[\overline{3}_{Q,\ell}]$.
We have 
    \begin{equation}
    \mathcal{Q}[3_{Q,\ell}(\overline{3}_{Q,\ell})]=
    \left(
    \begin{array}{ccc}
   \pm\frac{1}{2}\pm\frac{\beta}{2\sqrt{3}}
  +X[3_{Q,\ell}(\overline{3}_{Q,\ell})]
   & 0                                      & 0                         \\
    0                                      & \mp\frac{1}{2}\pm\frac{\beta}{2\sqrt{3}}+X[3_{Q,\ell}(\overline{3}_{Q,\ell})]
    & 0                         \\
    0                                      &                                        & \mp\frac{\beta}{\sqrt{3}}+
    X[3_{Q,\ell}(\overline{3}_{Q,\ell})]
    \\
    \end{array}
    \right),
    \label{charge3}
    \end{equation}        
The  upper (down) signs correspond to the $3$  ($\bar 3 $)  representations. The $X$ values for $3_{Q,\ell}$, and $\overline{3}_{Q,\ell}$ states are indicated, respectively, as $X[3_{Q,\ell}]$ and $X[\overline{3}_{Q,\ell}]$. We will use a similar notation $X[f]$ and $X[\overline{f} ]$ if, instead of referring to all fields in an entire multiplet, we refer to the specific field $f$. \\

The charges of the first two components of \eqref{assignment} and \eqref{assignment2}  must match  the charges of the corresponding SM fields. This matching fixes uniquely the $X$ values for the fermions as functions of $\beta$.
Let us consider  the case of the quark triplet $3_Q$. If we match the $\mbox{up}$ quark charge $1/2+\beta/(2\sqrt{3})+X[{3}_{Q}]=2/3$, we obtain $ X[{3}_{Q}]=1/6-\beta/(2\sqrt{3})$.  The correct $\mbox{down}$ quark charge follows automatically, and it does not represent an independent constraint.
In the same way we get  $X[{3}_{\ell}] =-1/2-\beta/(2\sqrt{3})$  in the lepton sector. \\
Let us now consider the entries of the matrix \eqref{charge3} for the antitriplets.  By repeating the same reasoning for the quark assignment in \eqref{assignment} we get  $X[\overline{3}_Q] =1/6+\beta/(2\sqrt{3})$, whereas  for the lepton one in \eqref{assignment2} we obtain $X[\overline{3}_\ell] =-1/2+\beta/(2\sqrt{3})$\footnote{We are using the same convention of  Ref.\,\cite{Buras:2012dp}, opposite to the one in Ref.\,\cite{Diaz:2004fs}.}. 
Negative values for $\beta$ can be related to the positive ones by taking the complex conjugate in the covariant derivative of each model, which in turn is equivalent to replace $3$ with $\bar 3$ in the fermion content of each particular model. 

The non-standard fields, 
$\psi_i, \Psi_j, \xi_k, \zeta_r$, have the same $X$ values of the other members of the respective (anti)triplet, because of gauge invariance.
Thus  using  eq.(\ref{charge3}) we deduce their charges,  $\mathcal{Q}[\psi_L]=1/6 - \sqrt{3}/2\,\beta$, $\mathcal{Q}[\Psi_L]=1/6 + \sqrt{3}/2\,\beta$,
$\mathcal{Q}[\xi_L]=-1/2 - \sqrt{3}/2\,\beta$ and $\mathcal{Q}[\zeta_L]=-1/2 + \sqrt{3}/2\,\beta$.

As far as right handed fields are concerned we assume, in analogy to the SM, that they are assigned to singlet representations of $SU_L(3)$.
Since the action of the charge operator $\hat{\mathcal{Q}}$ on the singlet 
 is given by 
$\hat{\mathcal{Q}} |1\rangle = X\, |1\rangle$, the $X$ value of a right-handed field coincides with its charge.  
In our notation, given a  left-handed fermion field $f_L$, belonging to a  triplet or anti-triplet $SU_L(3)$ representation, we indicate with $f_R$ the corresponding right-handed field in the $SU_L(3)$ singlet representation.
Let us observe that  the presence of
right-handed quark singlets is demanded for the model to be  $SU(3)_C$ vector-like. 
Table\,\eqref{tab:matter0} summarises the quantum numbers of the 331 fermions.

\begin{table}[tb!]
\centering
\begin{tabular}{|c|ccc|c|ccc|}
    \hline
    \hline
    & $SU_c(3)$ & $SU_L(3)$ & $U_X(1)$ &    & $SU_c(3)$ & $SU_L(3)$      & $U_X(1)  $    \\
    \hline
$\left(\begin{array}{c}u_L\\ d_L\\ \psi^{}_L\end{array}\right)_{i}$   & $3$       & $3$            & $\frac{1}{6}-\frac{\beta}{2\sqrt{3}}$ &
$\left(\begin{array}{c}\nu_L\\ \ell_L\\ \xi_L\end{array}\right)_{i}$                & $1$       & $3$ & $-\frac{1}{2}-\frac{\beta}{2\sqrt{3}}$ \\
$\left(\begin{array}{c}d_L\\ -u_L\\ \Psi_L\end{array}\right)_i$ & $3$ & $\overline{3}$ & $\frac{1}{6}+\frac{\beta}{2\sqrt{3}}$ &
$\left(\begin{array}{c}\ell_L\\ -\nu_L\\ \zeta_L\end{array}\right)_i$                & $1$       & $\overline{3}$ & $-\frac{1}{2}+\frac{\beta}{2\sqrt{3}}$ \\
$u_{R_i}$ & $3$ & $1$ & $\frac{2}{3}$ & & & & \\
$d_{R_i}$ & $3$ & $1$ & $-\frac{1}{3}$ & $\ell_{R_i}$ & $1$       & $1$            & $-1$ \\
$\psi_{R_i}$  & $3$ & $1$ & $\frac{1}{6}-\frac{\sqrt{3}\beta}{2}$ & $\xi_{R_i}$   & $1$ & $1$ & $-\frac{1}{2}-\frac{\sqrt{3}\beta}{2}$\\
$\Psi_{R_i}$  & $3$ & $1$ & $\frac{1}{6}+\frac{\sqrt{3}\beta}{2}$ & $\zeta_{R_i}$ & $1$ & $1$ & $-\frac{1}{2}+\frac{\sqrt{3}\beta}{2}$\\
    \hline
    \hline
\end{tabular}
\caption{\label{tab:matter0} Assigned representations and $X$ values of matter fields in terms of the parameter $\beta$ for a generic $331$ model. 
The subscript $i$ is a generic generation index. }
\end{table}

Let us now us consider the 331 gauge bosons.
Differently from the standard models we have 8  $W^a_\mu$ bosons for $SU_L(3)$ and one ${B_{X}}_{\mu}$ boson for $U_X(1)$. The covariant derivative acting on field representations are
\begin{itemize}
	\item triplet: $D_\mu=\partial_\mu-ig W^a_\mu T^a-ig_X X {B_{X}}_{\mu} T^9$;
	\item antitriplet: $D_\mu=\partial_\mu+ig W^a_\mu (T^a)^T-ig_X X {B_{X}}_{\mu} T^9$;
	\item singlet: $D_\mu=\partial_\mu-ig_X X {B_{X}}_{\mu} T^9$;
\end{itemize}
where the $g$ and $g_X$ are gauge couplings  and a sum over $a$ is understood.

We arrange the $SU(3)_\text L$ gauge bosons $W_\mu^a$ as
\begin{equation} \label{bosonmatrix}
	W_\mu=W_\mu^aT^a=\frac 1 2 \begin{pmatrix}
			W_\mu^3+\frac 1 {\sqrt 3} W_\mu^8  & \sqrt 2 W_\mu^+                     &  \sqrt 2 Y_\mu^{Q_Y} \\
			\sqrt 2 W_\mu^-                    &  -W_\mu^3+\frac 1 {\sqrt 3} W_\mu^8 & \sqrt 2 V_\mu^{Q_V}  \\
			\sqrt 2 Y_\mu^{-Q_Y}               & \sqrt 2 V_\mu^{-Q_V}                & -\frac 2 {\sqrt 3} W_\mu^8\end{pmatrix}
\end{equation}
where 
\begin{eqnarray}
    W^{\pm}_\mu     &=& \frac{1}{\sqrt 2} (W^1_\mu\mp i W_\mu^2)\nonumber\\
    V_\mu^{\pm Q_V} &=& \frac{1}{\sqrt 2} (W^6_\mu\mp i W_\mu^7)\\
    Y_\mu^{\pm Q_Y} &=& \frac 1 {\sqrt 2} (W^4_\mu\mp i W_\mu^5)\nonumber
    \label{gaugeboson}
    \end{eqnarray}
The action of the charge operator $\hat{\mathcal{Q}}$ on gauge bosons is given by the commutator 
$ \left[ \hat{\mathcal{Q}}, W_\mu\right]$, and it can be represented by the matrix     \begin{equation}
        \mathcal{Q}^{\rm GB} = \left(
        \begin{array}{ccc}
        0                                    & 1                                   &   \frac{1}{2} + \beta \frac{\sqrt{3}}{2}\\
        -1                                   & 0                                   & -\frac{1}{2} + \beta \frac{\sqrt{3}}{2} \\
        -\frac{1}{2}-\beta\frac{\sqrt{3}}{2} & \frac{1}{2}-\beta\frac{\sqrt{3}}{2} & 0 
        \end{array}
        \right) \label{chargematrix}
\end{equation}
where each entry represents the charge of the gauge boson in the corresponding entry of the matrix \ref{bosonmatrix}.
From eq.(\ref{chargematrix}) we see that there are:
\begin{enumerate}[label=(\roman*)]
\item three neutral gauge bosons,  $W^3_\mu$, $W^8_\mu$ and  ${B_{X}}_{\mu}$. After the spontaneous symmetry breakings, the neutral gauge bosons mix originating three mass eigenstates, the SM $\gamma$ and $Z^0$ plus an additional neutral BSM gauge boson, generally indicated with $Z^\prime$; 
\item  two bosons with unit charges, corresponding to $W^\pm$; 
\item four BSM bosons $Y_\mu^{\pm Q_Y}, V_\mu^{\pm Q_V}$, with charges $\pm \mathcal{Q}_Y=\pm\frac{1}{2}\pm \beta\frac{\sqrt{3}}{2}$ and $\pm \mathcal{Q}_V=\mp\frac{1}{2}\pm \beta\frac{\sqrt{3}}{2}$, that depend on the value of $\beta$. 
\end{enumerate}

It is useful at this point to summarize the electric charges of BSM fermions and BSM bosons: 
\begin{center}
\begin{tabular}{|c|c|c|c|c||c|c|c|}
\hline
field &$\psi$&$\Psi$ &$\xi$ &$\zeta$ &$\quad Z'\quad$ &$V $ &$Y $\\
\hline
$\mathcal{Q}$ & $\frac{1}{6}-\frac{\sqrt{3}}{2}\beta$ & $\frac{1}{6}+\frac{\sqrt{3}}{2}\beta$
& $-\frac{1}{2}-\frac{\sqrt{3}}{2}\beta$& $-\frac{1}{2}+\frac{\sqrt{3}}{2}\beta$
& 0& $\mp \frac{1}{2}\pm \frac{\sqrt{3}}{2}\beta$& $\pm \frac{1}{2}\pm \frac{\sqrt{3}}{2}\beta$\\
\hline
\end{tabular}
\end{center}

Depending on the choice of the parameter $\beta$ the electric charges of the  new fields can be  exotic. With exotic we mean charges different from $\pm1/3,\pm 2/3$ for extra quarks, and different from $\pm1$ for extra leptons and gauge bosons. We consider as a minimal requirement that the new gauge bosons have at least integer charge values, even if there are 331 models that discard this assumption, see for instance  \cite{RomeroAbad:2020zpv}.

By imposing that the new gauge bosons  $Y_\mu^{\pm \mathcal{Q}_Y} $ and $ V_\mu^{\pm \mathcal{Q}_V}$ have integer charge, it follows that the allowed values for the parameter $\beta$ are 
\begin{equation}\label{beta}
\beta = \pm\frac{2n+1}{\sqrt{3}}\equiv\frac{r}{\sqrt{3}} 
\end{equation}
where $r$ is an odd integer number (positive or negative). 
A few examples of $X$ and $\hat{\mathcal{Q}}$ charges for new fields  are reported in Table\,(\ref{Xcharge}), corresponding to
 $r=\pm 1, \pm 3,\pm 5$.
We immediately see that only the choice $r=\pm 1$  guarantees no exotic charges. It is necessary to warn that another constraint ($|r| \le 3$) comes from spontaneous symmetry breaking, as we will see in sect. \ref{Gaugesector}.
In Table\,(\ref{Xcharge})  we report  the case $\beta=\pm 5/\sqrt{3}$ only to remark that in the minimal assumptions made up to now
there is nothing that prevents $|r|$ from having values  larger than $3$.
 From Table\,(\ref{Xcharge}) it is also manifest  that the exchange  $r \longleftrightarrow -r$ is equivalent to trade triplets for anti-triplets and viceversa. \\

\begin{table}[ht!]
\begin{tabular}{|c|c|c|c|c||c|c|}
    \hline
$r$          &$+1$    &$-1$   & $+3$  &$-3$   & $+5$   & $-5$\\
\hline
$X[\psi]$    &$0$     &$1/3$  & $-1/3$&$2/3$  & $-2/3$ & $1$ \\
$X[\Psi]$    &$1/3$   &$0$    & $2/3$ &$-1/3$ & $1$    & $-2/3$\\
$X[\xi]$     &$-2/3$  &$-1/3$ & $-1$  &$0$    & $-4/3$ & $1/3$\\
$X(\zeta)$   &$-1/3$  &$-2/3$ & $0$   &$-1$   & $1/3$  & $-4/3$\\
\hline 
$\mathcal{Q}[\psi]$    &$-1/3$  &$2/3$  & $-4/3$&$5/3$  & $-7/3$ & $8/3$ \\
$\mathcal{Q}[\Psi]$    &$2/3$   &$-1/3$ & $5/3$ &$-4/3$ & $8/3$  & $-7/3$ \\
$\mathcal{Q}[\xi]$     &$-1$    &$0$    & $-2$  &$1$    & $-3$   & $2$\\
$\mathcal{Q}[\zeta]$   &$0$     &$-1$   & $1$   &$-2$   & $2$    & $-3$\\
\hline
$|\mathcal{Q}[V]|$     &$0$     &$1$    & $1$   &$2$    & $2$    & $3$\\
$|\mathcal{Q}[Y]|$     &$1$     &$0$    & $2$   &$1$    & $3$    & $2$\\
    \hline
\end{tabular}
\caption{\label{Xcharge} We list the $X$ values and the charges for the third components of quark triplet $3_Q$, lepton triplet $3_\ell$, quark anti-triplet $\overline{3}_Q$, lepton anti-triplet $\overline{3}_\ell$, and the  absolute values of the charges of BSM gauge bosons for sample  values of $r$ ($\beta = r/\sqrt{3}$). }
\label{table:charges}
\end{table}


The 331 models are fully characterized  by the choice  of the  electric charge embedding, namely the pair $(\beta,X)$,  of its fermionic  content and multiplet assignemnts, and of the scalar sector.
There are two major quantum field theory constraints, the absence of gauge anomalies and the possibility of QCD asymptotic freedom.

The fermionic structure  is restricted by requiring the cancellation of the triangular anomalies. Both $SU(3)_L$ and $U(1)_X$ are anomalous, contrary to the SM, where only $U(1)_Y$ is.
In the SM,
 gauge anomalies are cancelled out for every fermion family. This property is lost in 331 models and  correlation between families is needed. 
Ref.\,\cite{Diaz:2004fs} discusses in detail this topic for an arbitrary value of $\beta$ and number of generations.  
Below, we outline anomaly cancellation calculations assuming  $N_{Q (\ell)}^{(3)}$ and $N_{Q (\ell)}^{(\overline{3})}$ quark  (lepton) generations transforming as triplets and anti-triplets with the  structure outlined above. We underline that when we refer to lepton generations, we mean generations of active leptons, whose left-handed parts are embedded in triplets or antitriplets.
We define the total number of quark (lepton) generations as \begin{equation}
 N_{Q(\ell)}^F=N_{Q(\ell)}^{(3)}+N_{Q(\ell)}^{(\overline{3})} \label{eq:trpletsum}
 \end{equation}
The anomaly cancellations  with their implications read as

\begin{itemize}

\item[A1)] $[SU(3)_L]^3:$\\

$N_c N_Q^{(3)} - N_c N_Q^{(\overline{3})} + N_\ell^{(3)} - N_\ell^{(\overline{3})} = 0$\\

$\Rightarrow N_Q^{(3)}  =  \frac{N_c N_Q^F+N_\ell^F}{2N_c} - \frac{N_\ell^{(3)}}{N_c}$

\item[A2)] $[SU(3)_L]^2\otimes U(1)_X$:\\

$3 N_c \left[N_Q^{(3)} X[3_Q]+ N_Q^{(\overline{3})} X[\overline{3}_Q]\right]+ 3 N_\ell^{(3)} X[3_\ell]+ 3 N_\ell^{(\overline{3})}  X[\overline{3}_\ell] =0$\\

$\Rightarrow N_Q^F=\frac{3}{N_c}N_\ell^F$
\qquad \quad (assuming  A1))

\item[A3)] $[U(1)_X]^3$ \\

    $N_c \left[N_Q^{(3)} \left( 3 X[3_Q]^3 - X[\psi_R]^3\right)+ N_Q^{(\overline{3})} \left(3 X[\overline{3}_Q]^3 - X[\Psi_R]^3\right) - N_Q^F \left(X[u_R]^3 + X[d_R]^3\right) \right]+$\\
    $+ N_\ell^{(3)} \left(3 X[3_\ell]^3 - X[\xi_R]^3\right) + N_\ell^{(\overline{3})} \left(3 X[\overline{3}_\ell]^3 - X[\zeta_R]^3 \right) - N_\ell^F X[\ell_R]^3=0$\\

    $\Rightarrow$ No condition (always true).\\

\item[A4)] $[SU(3)_c]^2\otimes U(1)_X:$\\

$N_c \left[N_Q^{(3)} \left( 3 X[3_Q] - X[\psi_R]\right)+ N_Q^{(\overline{3})} \left(3 X[\overline{3}_Q] - X[\Psi_R]\right) - N_Q^F \left(X[u_R]+ X[d_R]\right) \right] =0$\\

 $\Rightarrow$ No condition (always true).\\

\item[A5)] $[\text{Grav}]^2\otimes U(1)_X $ \\

$N_c \left[N_Q^{(3)} \left( 3 X[3_Q] - X[\psi_R]\right)+ N_Q^{(\overline{3})} \left(3 X[\overline{3}_Q]-X[\Psi_R]\right) - N_Q^F \left(X[u_R]+ X[d_R]\right) \right] +$\\
    $+ N_\ell^{(3)} \left(3 X[3_\ell] - X[\xi_R]\right) + N_\ell^{(\overline{3})} \left(3 X[\overline{3}_\ell]-X[\zeta_R]\right) - N_\ell^F X[\ell_R]=0$\\

  $\Rightarrow$ No condition (always true).  

\end{itemize}
where $N_c$ refers to the color $SU_3(N_c)$,  always fixed to $N_c=3$. The minus signs in the sum are due to the different fields chirality. We remark that these relations hold even
under the common assumption $\zeta_L^a=(\ell^a_R)^c$.

Before drawing  conclusions from the anomaly cancellation, let us
 summarize the assumptions we have done on an otherwise completely general 331 model:
\begin{itemize} \item all matter multiplets include SM fermions, excluding the possibility of fermion multiplets including only BSM fields;
\item all gauge bosons, including BSM ones,  have integer charges;
\item all right-handed, $SU(3)$ singlet, fermions have a left-handed counterpart.
\end{itemize}
We underline that we could have 
assumed the presence of additional sterile right-handed or left-handed neutrinos in an arbitrary number without changing  the above A1)-A7) conditions. Nor it would have changed the definition of $N^F_{Q(\ell)}$, which refers to weak interacting fermions.

We found from relation A2) 
\begin{eqnarray}
N_\ell^F  & = & N_Q^F \equiv N_F\,, \label{eqs:familyconstr1} 
\end{eqnarray}
namely the number of quark and charged lepton families must be the same.
Combining A1) with A2) it follows 

\begin{eqnarray}
N_Q^{(3)} & = & \frac{1}{3} ( 2 N_F - N_\ell^{(3)}) \label{eqs:familyconstr2}\,. 
\end{eqnarray}
Both sides of the equality must be integers. 
This is  true if $N_F$ is a multiple of the color $N_c=3$, as in minimal models where $N_\ell^{(\bar3)}=0$. However such proportionality is not strictly necessary.
An example is  $N_F=4$ with $N_\ell^{(3)}=N_\ell^{(\bar{3})}=2$ that gives $N_Q^{(3)}=N_Q^{(\bar{3})}=2$ (see also Ref.\,\cite{Diaz:2004fs}). 

Another constraints is given by $SU(3)_c$ asymptotic freedom. As in SM QCD, perturbative calculations limit the number of quark flavours to be  equal to or less
than 16. Since each family in 331 models has three flavours, that yields a constraint on the number of families: $N_F\le 5$.

\vskip10.mm

Summarizing, under the minimal assumptions discussed above,  we can classify  331 models by means of $\beta= r/\sqrt 3$, where $r=\pm1,\pm 3$ 
(see for instance \cite{Pleitez:2021abk}), 
 the family number $N_F\le 5$, and the number of   triplets or antitriplets for either quarks or leptons.

Fixing  one of the 4 possible values for $\beta$, and assuming $N_c = N_F = 3$, we have two possible choices, namely $N_Q^3 =2, 1$, that we name  case A[$\beta$] and B[$\beta$], respectively. In Table \ref{tab:matter} we report the numbers of  quark and leptons in triplets and antitriplets according to the relations
 \eqref{eq:trpletsum},\eqref{eqs:familyconstr1}   and (\ref{eqs:familyconstr2}).
 Note that this classification   works only when   triplet and anti-triplets are used, but  it is also possible  to assign matter fields to different representations like sextects, see for instance Ref. \cite{Fonseca:2016tbn}.

\begin{table}[ht!]
\centering
\begin{tabular}{|c|c|c|c|c|}
    \hline
 $\qquad$ &  $N_Q^{(3)}$ & $N_Q^{(\bar{3})}$  & $N_\ell^{(3)}$ & $N_\ell^{(\bar{3})}$\\
\hline
A[$\beta$] & 2 & 1  & 0 & 3 \\
B[$\beta$] & 1 & 2  & 3 & 0 \\
    \hline
\end{tabular}
\caption{\label{tab:matter} Number of triplets and anti-triplets in the lepton and quark sectors in case A[$\beta$] and B[$\beta$] (see text). 
}
\end{table}

Models $A[\beta]$ and $B[-\beta]$ are related by the simultaneous exchange of fermion triplets $\to$ antitriplets and $\beta\to -\beta$. It implies that matter and gauge fields possess the same set of charges in both models \footnote{A complete symmetry between the two models requires additional conditions in the Higgs sector \cite{Hue:2018dqf}.}.
Several of these models have been discussed in literature: $A[1/\sqrt{3}]$ \,\cite{Montero:1992jk,Buras:2012dp, Boucenna:2014ela}, 
$A[-1/\sqrt{3}]$\,\cite{Cabarcas:2008ys, Profumo:2013sca, CarcamoHernandez:2013krw, Alves:2022hcp}, 
$A[\sqrt{3}]$ \,\cite{Singer:1980sw, Pisano:1992bxx, Foot:1992rh,  Montero:1998yw,Catano:2012kw, Frampton:1992wt, Correia:2017vxa,Corcella:2018eib}.
The case A$[\sqrt{3}]$ is known as the minimal 331 model and will be considered in this work.

We remark that in order to fully specify a model fixing the matter content as  above is not enough, but it has to be complemented by  the assignment of the Higgs structures and vacuum
alignments, see below.

\section{The model}\label{app:model}

\subsection{The scalar sector}  As mentioned in the introduction, the spontaneous symmetry breaking  (SSB) pattern of 331 models occurs in two steps: the first one allows to recover the SM group structure,  the second one is identifiable as the electroweak  SSB. The 331 models are intrinsically multi-Higgs models, which, as in the SM, are $SU(3)_C$ singlets acquiring non-vanishing
vacuum expectation values after SSB. They  are responsible of  mass terms for gauge bosons and fermions coupling to them.  
The requirement of gauge invariance of the Lagrangian term coupling  one Higgs  and two fermions constrains the Higgs representations.
Namely, since the fermions $\psi$ transform either as a $3$ or as a $\bar 3$ under $SU(3)_L$, we only have the following possibilities~\cite{Diaz:2003dk} for the Higgs field $\Phi$:
\begin{itemize}
	\item $\bar\psi_L\psi_R\Phi$, with  $\Phi\sim3$;
	\item $\bar\psi_L(\psi_L)^c\Phi$, with  $\Phi\sim6$;
	\item $\bar\psi_R(\psi_R)^c\Phi$, with  $\Phi\sim1$;
	\item $\bar\psi_L(\psi'_L)^c\Phi$, with $\Phi\sim\bar3$;
\end{itemize}
The last possibility is allowed only if $\psi'_L$ are  completely exotic triplets with charged particles which are right handed charge-conjugated components \cite{Descotes-Genon:2017ptp}, otherwise is  zero.
Singlets scalars are also scalars under $U(1)_X$ because of electromagnetic invariance. Thus, after the two steps of SSB, their
vacuum expectation value will never give rise to a mass
term for the gauge bosons or the charged fermions. Sextets  are needed in some models, for instance to give  Majorana masses to neutrinos as in Ref. \cite{Foot:1992rh} or to allow  a doubly-charged scalar to decay into same-sign leptons \cite{Corcella:2018eib}. Both singlets and sextets are not needed for the model and  the phenomenology we are investigating, so we limit to the simpler, and widely used in literature, assumption of  SSB  by triplets only. 
The $X$ values of the triplets, and therefore also the charges of their components, are constrained by the  requirement that the above Higgs-fermion coupling terms in the Lagrangian are also invariant under $U(1)_X$. 

The Higgs triplets are:
\begin{equation}
\chi=
\begin{pmatrix} 
\chi^{++} \\ \chi^+ \\ \chi^0
\end{pmatrix}, \hspace{1cm}
\rho=
\begin{pmatrix} 
\rho^{+} \\ \rho^0 \\ \rho^-
\end{pmatrix},\,\hspace{1cm}
\eta=
\begin{pmatrix} 
\eta^{0} \\ \eta^- \\ \eta^{--}
\end{pmatrix}.
\label{eq:higgstriplet}
\end{equation}
Their $X$ values are listed in Table \ref{tab:fermion}.
The neutral components acquire a vacuum expectation value, namely $\langle\chi^0\rangle \equiv u $, $\langle\rho^0\rangle \equiv v $, $\langle\eta^0\rangle \equiv v' $. 
 In the first SSB, triggered by the triplet $\chi$, there are five gauge fields that acquire mass.
In the second SSB, which requires the pair of triplets $\rho$ and $\eta$, three gauge fields acquire mass, leaving  massless the field associated to the unbroken charge generator, that is the photon.  

\subsection{The gauge sector}
\label{app:Gaugesector}

Boson masses stem from the expansion of the covariant derivatives of the scalar sector
\begin{equation}
    \mathcal{L}_{S} = (D_\mu\chi)(D^\mu \chi)^\dagger + (D_\mu\rho)(D^\mu \rho)^\dagger + (D_\mu\eta)(D^\mu \eta)^\dagger 
\end{equation}
The neutral gauge bosons matrix is not diagonal. In the base $W_\mu^3,\,W_\mu^8,\, B_{X\,\mu}$, it is given by

{\footnotesize
\begin{equation}
    M_0 =\frac{1}{4} \left(
    \begin{array}{ccc}
    g^2 v^2_+                                                      &  \frac{g^2}{\sqrt{3}} v_-^2                                                              & - \frac{gg_X}{3\sqrt{2}} \left(v_+^2\sqrt{3}-v_-^2\beta\right)\\
    \frac{g^2}{\sqrt{3}} v_-^2                                     &  \frac{g^2}{3} \left(4 u^2 + v_+^2\right)                                                & -\frac{gg_X}{9\sqrt{2}} \left(u^2 4\sqrt{3}\beta + v_+^2\sqrt{3}\beta + 3 v_-^2 \right)\\
    - \frac{gg_X}{3\sqrt{2}} \left(v_+^2\sqrt{3}-v_-^2\beta\right) & -\frac{gg_X}{9\sqrt{2}} \left(u^2 4\sqrt{3}\beta + v_+^2\sqrt{3}\beta + 3 v_-^2 \right) & \frac{g_X^2}{18}\left(4 u^2 \beta^2 + v_+^2(3+\beta^2) + v_-^2 2\sqrt{3}\right)
    \end{array}
    \right)
\end{equation}}
where $v^2_- =v^{\prime\,2}- v^2$. \\
Since $u^2\gg v_+^2,\,v_-^2$ we can diagonalize this matrix in two steps. In the first one $B_{X\,\mu},\,W_\mu^8$ mix giving rise to $B'_\mu$ and $Z'_\mu$ as  
\begin{eqnarray}
    \left(\begin{array}{c} W^8_{\mu}\\ B_{X\,\mu}\end{array}\right) \equiv
    \left(\begin{array}{cc}
     \cos\theta_{331} &-\sin\theta_{331} \\
        \sin\theta_{331}  &\cos\theta_{331}                     \\
    \end{array}\right)
    \left(\begin{array}{c} B'_\mu\\ Z'_\mu\end{array}\right)\,.
    \label{eq:theta_331}
\end{eqnarray}
With the above definition of the mixing angle, by diagonalizing $M_0$ we obtain 
\begin{eqnarray}
 &&\sin \theta_{331}\simeq \frac{g}{\sqrt{g^2 + g_X^2 \beta^2/6}}\,.\label{eq331}
\end{eqnarray}
Then at the second step $B'_\mu, Z_\mu^\prime$ mix both with $W_\mu^3$ but the mixing between 
$Z_\mu^\prime$ and $W_\mu^3$ is small compared to the one between $B'_\mu$ and $W_\mu^3$, therefore the first one can be neglected and we have
\begin{eqnarray}
    \left(\begin{array}{c} W_\mu^3\\ B'_{X\,\mu}\end{array}\right) \equiv
    \left(\begin{array}{cc}
    \cos\theta_W                  &  
    \sin\theta_W                  \\
    -\sin\theta_W & \cos\theta_W\\
    \end{array}\right)
    \left(\begin{array}{c} Z_\mu\\ A_\mu\end{array}\right)\label{eqw}
\end{eqnarray}
where $A_\mu$ is the Standard Model photon field. 
The corresponding boson masses  are
\begin{eqnarray}
m_A&=& 0\,,\\
m_Z^2          &   \simeq  & 
\frac{g^2}{4\cos^2\theta_W}v_{\rm SM}^2\,\\        
m_{Z^\prime}^2 &   \simeq   & 
\frac{g^2 u^2 \cos^2\theta_W}{3[1-(1+\beta^2)\sin^2\theta_W]}\,.\label{Zprime_mass}
\end{eqnarray}
We note that $m_W^2/m_Z^2 =\cos^2\theta_W$ and therefore we can identify $\theta_W$ with the Weinberg angle.

An important constraint for 331 models is obtained by imposing a matching condition for the coupling of $Z$ boson with fermions. In particular, considering the right-handed leptons $\ell_R$ with $X$-charge $X[\ell_R]=-1$, the covariant derivative is given by 
\begin{equation}
    D_\mu \ell_R =\partial_\mu \ell_R + i g_X \frac{B_{X_\mu}}{\sqrt{6}}\,\ell_R
\end{equation}
From Eqs.\,\eqref{eq:theta_331} and\,\eqref{eqw} we have that 
\begin{equation}
    B_{X_\mu}=\cos\theta_{331}\,Z'_\mu-\sin\theta_{331}\sin\theta_{W} Z_\mu
    +\sin\theta_{331}\cos\theta_{W} A_\mu
\end{equation}
Then by singling out the coupling with $Z$ boson and by matching with the corresponding Standard Model coupling $g \sin^2\theta_W/\cos\theta_W$ we get
\begin{equation}
    \frac{g_X}{g} =\frac{\sqrt{6} \sin\theta_W}{\cos\theta_W\sin\theta_{331}}=
    \frac{\sqrt{6} \sin\theta_W}{\sqrt{1-\sin^2\theta_W (1+\beta^2)}}\label{eqgxg}
\end{equation}
where we have used eq.\,(\ref{eq331}).
 Beyond tree level, the precise definition of $\sin^2\theta_W$ is renormalization scheme and
energy dependent. Extending the tree-level formula  $m_W^2/m_Z^2 =\cos^2\theta_W$ to all orders of
perturbation theory  results in the on-shell value of $\sin^2\theta_W$ \cite{Sirlin:1980nh}.
Another
widely used convention is the modified minimal subtraction ($\overline{\mathrm{MS}}$) scheme.
We can see that the tree level Eq. \eqref{eqgxg} exhibits a Landau pole when
$\sin^2\theta_W=1/(1+\beta^2)$. As in other 
non asymptotically free theories,
at the Landau pole the coupling constant $g_X$ becomes infinite. The energy scale of the Landau pole depends on  the running of $\sin^2\theta_W$, which in turn depends on 
the matter content of the model.
Besides, in the on-shell scheme, radiative corrections for $\sin^2\theta_W$ include $m_t^2$-dependent terms which can cause large corrections in higher orders. $\sin^2\theta_W$  depends not only on the gauge couplings, but also on the spontaneous symmetry breaking.
In the SM framework, using the $\overline{\mathrm{MS}}$ scheme, it  has been estimated that values of $ \sin^2\theta_W $ around the upper limit of the constraint $ \sin^2\theta_W < 0.25 $ do not arise before $\approx$ 4 TeV
\cite{Amoroso:2023uux}. There are 1-loop analyses that estimate that at large values of the coupling, that is even before the Landau pole is reached, the  model
 may lose its perturbative character \cite{Martinez:2006gb,Dias:2004dc,Dias:2009au,Dias:2005xj} (see also the discussion in \cite{Buras:2013dea}). At the same time, some of these analyses indicate how to push  the  non-perturbative regime, and the Landau pole, up to much higher scales,  including for example heavy resonances as active degrees of freedom in the running  \cite{Martinez:2006gb,Dias:2004wk}.
 We stress anyway that the typical energy scale of bilepton-pair production we consider is  of the order of $\approx 2\, m_Y\approx 2$ TeV. Given the considerations above, we assume that no significative barrier is posed by the Landau pole to the perturbative
analysis carried out in the present study.

We also note that the left-hand side of eq.\,(\ref{eqgxg})
is real if (see also \cite{Fonseca:2016tbn})
\begin{equation}
    |\beta|<\frac{\cos\theta_W}{\sin\theta_W}\simeq 1.83 \,.\label{eqbeta}
\end{equation}
This upper limit of $|\beta|$ is extremely important for the model building. As shown above the only allowed values for $|\beta|$ are $r/\sqrt{3}$ with $r$ integer. The constraint coming from eq.\,(\ref{eqbeta}) yields $n\leq 3$ and therefore models with $n>3$ are not allowed in this minimal framework.

\section{Decay rates and vertices}

\noindent
Here we report the decay widths of the VLQs introduced in the model:
\begin{eqnarray}
&&\Gamma(T \to Y^{++} d^a)  = \frac{g^2 |{{\cal V}_L}_{3a}|^2}{64 \pi}\frac{m^6_T -3 m^2_T m^4_Y+2m^6_Y }{m^3_T m^2_Y}\,,\\
&& \Gamma(D^i \to Y^{--} u^a)= \frac{g^2 |{{\cal U}_L}_{ia}|^2}{64 \pi}\frac{m^6_D -3 m^2_D m^4_Y+2m^6_Y }{m^3_D m^2_Y}\,,\\
&&\Gamma(T \to V^{+} u^a)  = \frac{g^2 |{{\cal U}_L}_{3a}|^2}{64 \pi}\frac{m^6_T -3 m^2_T m^4_V+2m^6_V }{m^3_T m^2_V}\,,\\
&& \Gamma(D^i \to V^{-} d^a)= \frac{g^2 |{{\cal V}_L}_{ia}|^2}{64 \pi}\frac{m^6_D -3 m^2_D m^4_V+2m^6_V }{m^3_D m^2_V}\,,
\end{eqnarray}
where $m_X$ stays for the mass of the $X$ field.
The bilepton $Y$ decay width is 
\begin{equation}
    \Gamma(Y \to \ell^\pm \ell^\pm)=\frac{g^2}{12 \pi} m_Y \, .
\end{equation}

\noindent
In the following tables we give the relevant vertices of the model:

\begin{longtable}[ht!]{cc|cc }
     \toprule
\multicolumn{4}{c}{ Gauge boson-fermion interactions}
\\
     \hline
      Vertex & Coupling & Vertex & Coupling \\ 
      \hline
      \multicolumn{2}{c|}{$V$ boson}&\multicolumn{2}{c}{$Y$ boson}\\ 
      \hline
	\includegraphics[height=2.2cm]{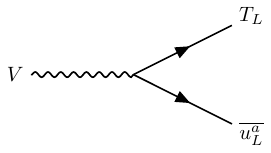}         & $\frac{g}{\sqrt{2}} \mathcal{U}_{L\,3a}$ &
	\includegraphics[height=2.2cm]{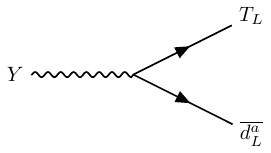}         & $-\frac{g}{\sqrt{2}} \mathcal{V}_{L\,3a}$ \\
	\hline
	\includegraphics[height=2.2cm]{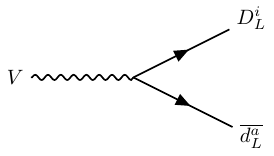}        & $\frac{g}{\sqrt{2}} \mathcal{V}_{L\,ia}$ &
	\includegraphics[height=2.2cm]{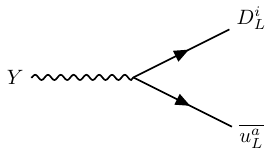}        &  $\frac{g}{\sqrt{2}} \mathcal{U}_{L\,ia}$ \\
	\hline
	\includegraphics[height=2.2cm]{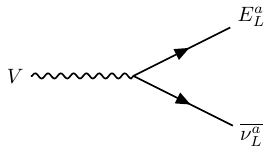}      & $\frac{g}{\sqrt{2}}$                   &
     	\includegraphics[height=2.2cm]{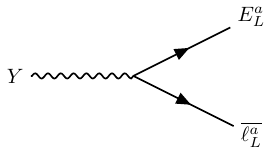}   & $-\frac{g}{\sqrt{2}}$            \\ 
      \hline
      \multicolumn{4}{c}{ $Z$ boson}\\ 
      \hline
      \includegraphics[height=2.2cm]{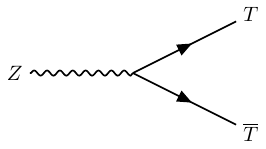}        & $\frac{g}{c_{w}} \left(\frac{1}{6} +\frac{\sqrt{3}}{2}\beta\right)s_{w}^{2}$ & 
      \includegraphics[height=2.2cm]{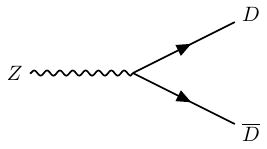}       & $\frac{g}{c_{w}} \left(-\frac{1}{6} +\frac{\sqrt{3}}{2}\beta\right)s_{w}^{2}$ \\
      \hline
      \includegraphics[height=2.2cm]{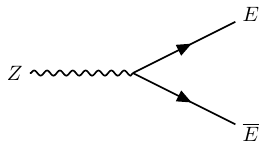}        & $\frac{g}{2c_{w}} \left(1-\sqrt{3}\beta\right)s_{w}^{2}$&
      &\\
      \hline
      \multicolumn{4}{c}{ $Z^\prime$ boson}\\ 
      \hline
    \includegraphics[height=2.2cm]{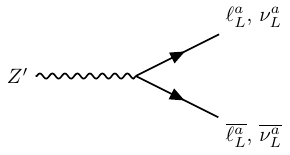}       &  $\frac{g\left[1-(1+\sqrt{3}\beta)s_{w}^{2}\right]}{2\sqrt{3}c_{w}\sqrt{1-(1+\beta^{2})s_{w}^{2}}}$ &
      \includegraphics[height=2.2cm]{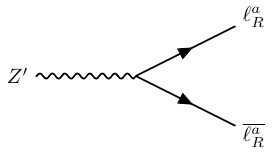}      & $-\frac{g\beta s_{w}^{2}}{c_{w}\sqrt{1-(1+\beta^{2})s_{w}^{2}}}$ \\
     \hline
     \includegraphics[height=2.2cm]{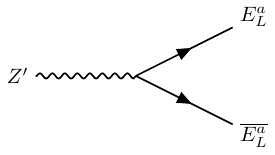}    &  $-\frac{g\left[2-\left(2-\sqrt{3}\beta (1-\sqrt{3}\beta)\right)s_{w}^{2}\right]}{2\sqrt{3}c_{w}\sqrt{1-(1+\beta^{2})s_{w}^{2}}}$ &
     \includegraphics[height=2.2cm]{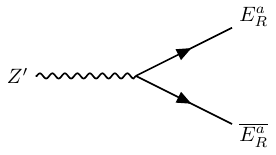}      & $-\frac{g(1-\sqrt{3}\beta)\beta s_{w}^{2}}{2c_{w}\sqrt{1-(1+\beta^{2})s_{w}^{2}}}$ \\
     \hline
     \includegraphics[height=2.2cm]{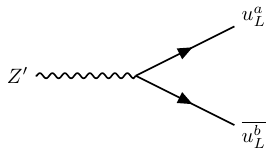}       &  $\frac{g\left[[-1+(\beta/\sqrt{3})s_{w}^{2}]\delta_{ab}+2c_{w}^{2}\mathcal{U}^{*}_{L\,3a}\mathcal{U}_{L\,3b}\right]}{2\sqrt{3}c_{w}\sqrt{1-(1+\beta^{2})s_{w}^{2}}}$ &
     \includegraphics[height=2.2cm]{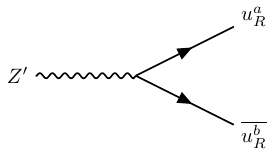}      &  $\frac{2g\beta s_{w}^{2}\delta_{ab}}{3c_{w}\sqrt{1-(1+\beta^{2})s_{w}^{2}}}$  \\
     \hline
     \includegraphics[height=2.2cm]{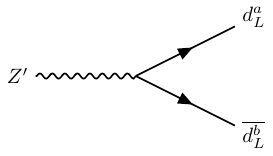}       &  $\frac{g\left[[-1+(\beta/\sqrt{3})s_{w}^{2}]\delta_{ab}+2c_{w}^{2}\mathcal{V}^{*}_{L\,3a}\mathcal{V}_{L\,3b}\right]}{2\sqrt{3}c_{w}\sqrt{1-(1+\beta^{2})s_{w}^{2}}}$ &
     \includegraphics[height=2.2cm]{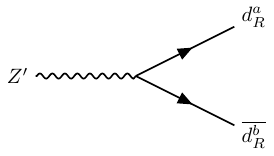}      &  $-\frac{g\beta s_{w}^{2}\delta_{ab}}{3c_{w}\sqrt{1-(1+\beta^{2})s_{w}^{2}}}$ \\
     \hline
     \includegraphics[height=2.2cm]{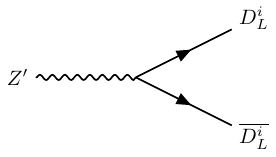}       &  $\frac{g \left[2+ \left(-2\sqrt{3}+\beta(1-3\sqrt{3}\beta)\right)s_{w}^{2}/\sqrt{3}\right]}{2\sqrt{3}c_{w}\sqrt{1-(1+\beta^{2})s_{w}^{2}}}$ &
     \includegraphics[height=2.2cm]{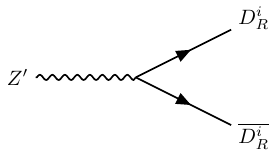}      &  $\frac{g \beta(1-3\sqrt{3}\beta)s_{w}^{2}}{6 c_{w}\sqrt{1-(1+\beta^{2})s_{w}^{2}}}$ \\
     \hline
     \includegraphics[height=2.2cm]{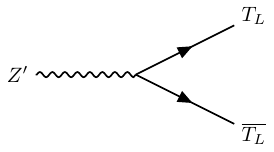}       &  $\frac{g \left[-2+ \left(2\sqrt{3}+\beta(1-3\sqrt{3}\beta)\right)s_{w}^{2}/\sqrt{3}\right]}{2\sqrt{3}c_{w}\sqrt{1-(1+\beta^{2})s_{w}^{2}}}$ &
     \includegraphics[height=2.2cm]{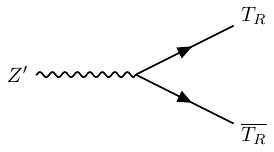}      &  $\frac{g \beta(1+3\sqrt{3}\beta)s_{w}^{2}}{6 c_{w}\sqrt{1-(1+\beta^{2})s_{w}^{2}}}$ \\

      \hline  
\end{longtable}

\end{document}